\documentclass[12pt,preprint]{aastex}

\def\ea{{\it et al.} \,}

\def\s-z{S-Z}

\def\boom{{\sc Boomerang} }
\def\boomn{{\sc Boomerang}}

\def\aa{{\it Astron. Astrophys.} \,}
\def\apj{{\it Ap. J.}\,}

\def\apjl{{\it Ap. J. Lett.}\,}

\def\nat{{\it Nature}\,}

\def\lta{\mathrel{\spose{\lower 3pt\hbox{$\mathchar"218$}}
     \raise 2.0pt\hbox{$\mathchar"13C$}}}
\def\gta{\mathrel{\spose{\lower 3pt\hbox{$\mathchar"218$}}
     \raisemodels do fit 2.0pt\hbox{$\mathchar"13E$}}}
\def\ge{\mathrel{\spose{\lower 3pt\hbox{$-$}}
     \raise 2.0pt\hbox{$\mathchar"13E$}}}
\def\le{\mathrel{\spose{\lower 3pt\hbox{$-$}}
     \raise 2.0pt\hbox{$\mathchar"13C$}}}

\def\2pr{^{\prime\prime}}

\def\greatsim{\mathrel{\raise.3ex\hbox{$>$\kern-.75em\lower1ex\hbox{$\sim$}}}}
\def\lesssim{\mathrel{\raise.3ex\hbox{$<$\kern-.75em\lower1ex\hbox{$\sim$}}}}

\begin{document}

\title{The \boom  North America instrument: a balloon-borne bolometric 
radiometer optimized for measurements of cosmic background radiation 
anisotropies from 0.3$^\circ$ to 4$^\circ$.}

\author{
F.~Piacentini\altaffilmark{1},
P.A.R.~Ade\altaffilmark{2},
R.~Bathia\altaffilmark{3},
J.J.~Bock\altaffilmark{3,4},
A.~Boscaleri\altaffilmark{5},
P.~Cardoni\altaffilmark{6},
B.P.~Crill\altaffilmark{3},
P.~de Bernardis\altaffilmark{1},
H.~Del~Castillo\altaffilmark{4},
G.~De~Troia\altaffilmark{1},
P.~Farese\altaffilmark{1},
M.~Giacometti\altaffilmark{1},
E.F.~Hivon\altaffilmark{3},
V.V.~Hristov\altaffilmark{3},
A.~Iacoangeli\altaffilmark{1},
A.E.~Lange\altaffilmark{3},
S.~Masi\altaffilmark{1},
P.D.~Mauskopf\altaffilmark{3,7},
L.~Miglio\altaffilmark{1,8},
C.B.~Netterfield\altaffilmark{3,8},
P.~Palangio\altaffilmark{9},
E.~Pascale\altaffilmark{3,5},
A.~Raccanelli\altaffilmark{1},
S.~Rao\altaffilmark{1,9},
G.~Romeo\altaffilmark{9},
J.~Ruhl\altaffilmark{10},
F.~Scaramuzzi\altaffilmark{6}}

\altaffiltext{1}{Dipartimento di Fisica, Universita' 
La Sapienza, P.le A. Moro 2, 00185, Roma, Italy, e-mail:
francesco.piacentini@roma1.infn.it}
\altaffiltext{2}{Department of Physics, Queen Mary and Westfield
College, Mile End Road, London, E1 4NS, U.K.}
\altaffiltext{3}{Department of Physics, Math, and Astronomy, California
Institute of Technology, Pasadena, CA, USA}
\altaffiltext{4}{Jet Propulsion Lab, Pasadena, CA, USA}
\altaffiltext{5}{IROE-CNR, Via Panciatichi 54, 50127 Firenze, Italy}
\altaffiltext{6}{Ente Nazionale Energie Alternative, Frascati, Italy}
\altaffiltext{7}{Dept. of Physics and Astronomy, University of Massachusets, 
Amherst, MA, USA}
\altaffiltext{8}{Department of Astronomy, University of Toronto, Canada }
\altaffiltext{9}{Istituto Nazionale di Geofisica, Roma, Italy}
\altaffiltext{10}{Dept. of Physics, Univ. of California, Santa Barbara, CA, 
USA}

\begin{abstract}
We describe the \boom North America (BNA) instrument, 
a balloon-borne bolometric radiometer designed to map 
the Cosmic Microwave Background (CMB) radiation with 
0.3$^\circ$ resolution over a significant portion of 
the sky.  This receiver employs new technologies in
bolometers, readout electronics, 
millimeter-wave optics and filters, cryogenics, scan and attitude 
reconstruction. All these subsystems are described in
detail in this paper. The system has been fully calibrated
in flight using a variety of techniques which are described
and compared.  Using this system, we have obtained a measurement of 
the first peak in the CMB angular power spectrum in a 
single, few hour long balloon flight.  The instrument described here
was a prototype of the \boom Long Duration Balloon (BLDB) 
experiment.
\end{abstract}


\section{Introduction}
\label{bintro}

The existence of the 2.7~K Cosmic Microwave Background (CMB) Radiation 
is evidence that our universe originated from a hot, dense plasma \cite{hu}.
This radiation was emitted
in the early universe, when the plasma cooled enough for
protons and electrons to form hydrogen atoms. Before 
recombination, photons and baryons where tightly coupled by
Compton scattering.
After recombination the photons and baryons are highly decoupled,
with a mean free path longer than the causal horizon.
This event is dated $\sim300000$ years after the Big Bang when the
universe was $\sim 1000$ times smaller and $\sim50000$ times 
younger (red shift z=1000). 
The properties of the CMB reflect the conditions of the early universe
and are closely linked with global properties of
the universe, such as the energy density ($\Omega_{tot}$), composition 
($\Omega_{\rm b}$, $\Omega_\Lambda$, $\Omega_{\rm DM}$), and expansion rate
(${\rm H}_0$).

In particular, variations in the brightness of the
CMB, or anisotropies, reflect variations in temperature, density, and
velocity of the last scattering surface. 
These fluctuations, that
originated from random noise at an earlier phase of expansion, are
the seeds for the formation of structures such as galaxies and clusters
of galaxies present in the universe today.
Precise measurements of the angular power spectrum of CMB
anisotropies \cite{WSS} will discriminate between competing cosmological 
models and, if the inflationary scenario is correct, will accurately
determine many of the physical parameters of the universe.

The COBE-DMR detection \cite{Smoot,DMRI}
of anisotropies in the CMB at large angular scales
provides a point of reference for theoretical models of the origin
of fluctuations in the CMB \cite{cmbbond}.  
Measurements at smaller angular scales are
needed to fully understand the nature of these fluctuations.
Since the launching of COBE, many other experiments have
made significant detections of CMB anisotropy at a wide
variety of angular scales, from $0.3^{\circ}$ to $10^{\circ}$,
but until 1998, none of the measurements provided enough information
for serious cosmological parameter determination.

The new generation of CMB experiments is designed to probe
models of structure formation with
a combination of higher sensitivity, sky coverage and resolving power.
Advances in detector technology have resulted in radiometers that
are over 100 times more sensitive than the COBE-DMR per unit time.  
In addition, improved
techniques for removing noise from atmospheric fluctuations with both
single dish instruments and interferometers allow
increased sensitivity with ground-based telescopes.  Long Duration
Balloon (LDB) platforms provide the opportunity to obtain the long integration
times needed for large sky coverage with balloon-borne telescopes.  

\boom is an experiment designed to measure the detailed structure
of the CMB at angular scales from
0.2$^{\circ}$ to $4^{\circ}$ with high sensitivity.  The \boom
instrument consists of a 1.3~m balloon-borne telescope and pointing
platform with a 300~mK bolometric array receiver.  The receiver is
contained inside a liquid Nitrogen (LN2) and liquid Helium 
(LHe) cryostat with a hold-time of two weeks.

The measurement technique consists of measuring a sky brightness map
by slowly scanning the telescope (and the full payload) in azimuth,
using the earth rotation to cover a wide sky region. There is no
mechanical chopper. The scan converts
CMB anisotropies at different angular scales into
detector signals at different sub-audio frequencies. The
instrument features a new total power readout of the detectors,
optimized to preserve the information content of the signal
while rejecting very low and very high frequency noise.
This approach pioneers in several aspects the HFI instrument
on the Planck satellite.

In this paper we describe the instrument prototype as it was 
used for the test flight, on Aug.30, 1997, to qualify 
all the flight subsystems for later use in the Antarctic 
LDB flight.
During the 1997 test flight we observed about 200 square degrees of sky
at high Galactic latitudes. Science results from that flight are reported
in \cite{b97mausk} and \cite{b97melch}. 
Here we give all the technical details of the instrument as well 
as its performance during the test flight: thermal performance,
bolometer loading and noise, scan performance, attitude reconstruction, 
calibration on Jupiter and on the CMB Dipole.
The LDB instrument is described in \cite{crill}.

\section{Instrument}
\label{instr}

\boom is designed to take advantage of the long integration time
possible from a balloon borne platform flown over the Antarctic.
Antarctic summer ballooning is very attractive for CMB anisotropy experiments
for two reasons. The first is that the flight duration 
(up to $\sim 20$ days)
allows for substantial sky coverage and deep checks for systematics; 
the second is that very low
foreground regions are observable in the direction generally opposite 
the sun during the Antarctic summer.

There are, however, a number of challenges peculiar to Antarctic
ballooning. The long flight duration requires special cryogenic systems.
The cosmic ray flux in polar regions is enhanced by a factor
of about ten with respect to North American latitudes, thus resulting in 
a high noise in standard bolometric detectors. 
The continual presence of the sun during an Antarctic LDB flight
is a general concern for the thermal performance of the payload and for
pickup in the sidelobes of the telescope.  The balloon is far from the 
ground equipment, so special data collection and telemetry systems have to 
be used, and interactivity with the system is reduced.

We describe in the following the solutions we have adopted to 
overcome these problems, developing custom subsystems:
a long duration cryostat, spider-web bolometers,
low sidelobes off-axis optics, special sun shields 
and telescope baffling, total power readout.

\begin{figure}[ht]
\epsscale{.6}
\begin{center}
\rotatebox{90}{
\plotone{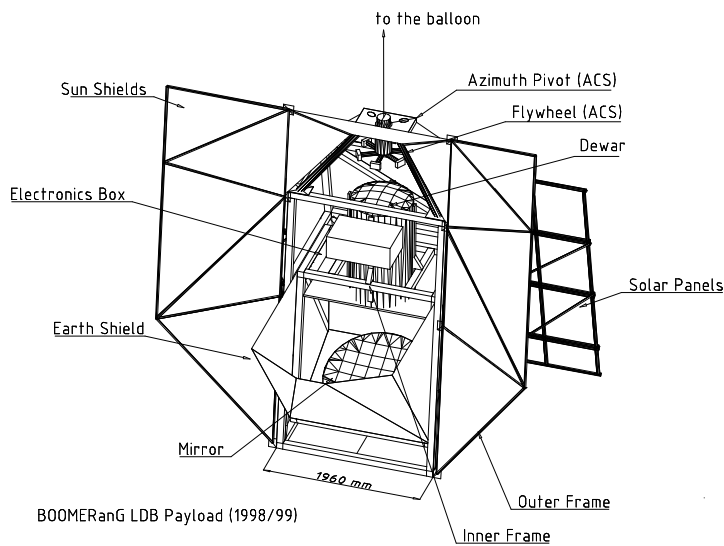}}
\caption{
\boom payload. Special shields in aluminized mylar protect from sun
and earth radiation reducing sidelobes. Radiation coming
from the sky is reflected into the cryostat by means of the 1.3~m primary
mirror. Solar panel are used for power supply in the LDB flight while
Lithium batteries ware used in the BNA flight.
The Attitude Contol System (ACS) provides pointing
and scanning of the telescope.
\label{fig:gondola}}
\end{center}
\end{figure}

A general view of the experiment with its main subsystems is 
shown in Figure~\ref{fig:gondola}.

\subsection{Optics}
\label{optics}

The \boom telescope consists of an ambient temperature 1.3~m diameter 
off-axis parabolic primary mirror (f=1280~mm, 45$^\circ$ off-axis)
which feeds cold reimaging secondary and tertiary mirrors
inside a large liquid helium cryostat.  The telescope and the cryostat are 
both mounted on an aluminum frame (the inner frame of the payload) 
which can be tipped in elevation by $-12^{\circ}$ to $+20^{\circ}$
to cover elevation angles from $33^{\circ}$ to $65^{\circ}$.
The primary mirror has a $45^{\circ}$ off-axis angle.
Radiation from the sky is reflected by the primary mirror and enters
the cryostat through a thin (50 $\mu$m) polypropylene window near
the prime focus.  Two circular windows side by side, each 66 mm in diameter,
are used.  This geometry provides a wide field of
view while allowing the use of thin window material which minimizes
the emission from this ambient temperature surface.  Filters rejecting
high frequency radiation are mounted on the 77~K and 2~K shields in
front of the cold reimaging optics.  Fast off-axis secondary and tertiary
mirrors surrounded by absorbing baffles reimage the prime focus onto the
detector focal plane. 

\begin{figure}[ht]
\epsscale{.55}
\begin{center}
\plotone{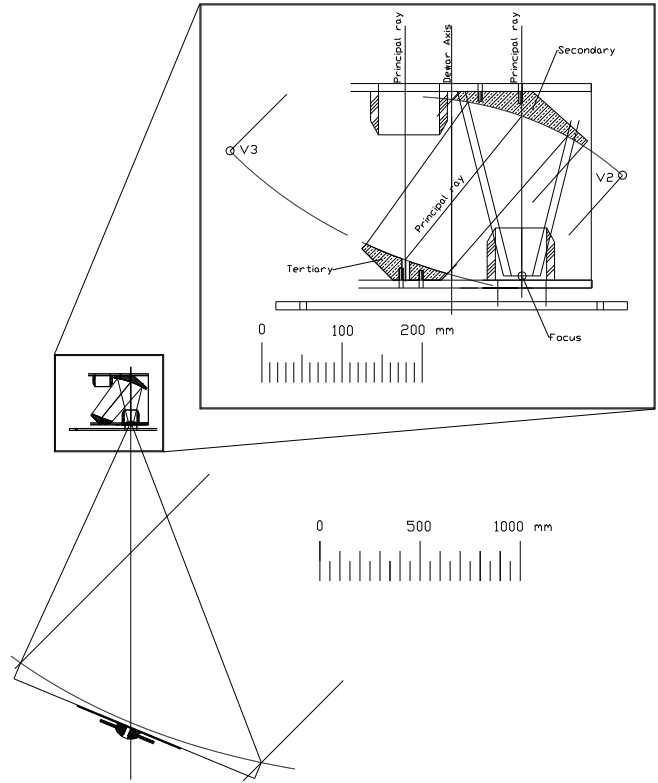}
\caption{
\boom optics. Secondary and tertiary mirrors  project the image on
the focal plane. They are cooled to 2~K inside the main cryostat.
An image of the 1.3~m primary mirror is formed on the tertiary
mirror which works as a cold Lyot stop, improving the off-axis
rejection of the photometer.
\label{fig:cold_o}}   
\end{center}
\end{figure}

The \boom optics (Figure~\ref{fig:cold_o}) is optimized for an 
array with widely separated
pixels.  The advantage of having large spacing between pixels in the
focal plane is the ability to difference (or compare) the signals 
from two such
pixels and remove correlated optical fluctuations such as temperature
drift of the telescope, while retaining high sensitivity to structure
on the sky at angles up to the pixel separation.  This scheme eliminates
the need for moving optical components and simplifies the design and
operation of the experiment.

We optimized the optics for diffraction limited performance at 1~mm over a
$2^{\circ} \times 5^{\circ}$ field of view.  The reimaging optics
are configured to form an image of the primary mirror at the
10~cm diameter tertiary mirror.  The size of the tertiary mirror therefore
limits the illumination pattern on the primary mirror, which is underfilled
by 50\% in area (85~cm in diameter) to improve sidelobe rejection.

The secondary mirror is an ellipsoid and the tertiary is a paraboloid,
10~cm in diameter, corresponding to an 85~cm diameter aperture on the 1.3~m
primary mirror. The equation describing the three mirrors is
\begin{equation}\label{eqn:opt}
z(r)=\frac{r^2}{R \left[ 1+\sqrt{1-(1+k)\frac{r^2}{R^2}} \right]}+Ar^4+Br^6
\end{equation}
with parameters $R$, $k$, $A$, $B$ as in Table~\ref{tab:mirrors}.

\begin{table}[ht]
\begin{center}
\begin{tabular}{lcccc}
\tableline
\tableline
Mirror & $R$ (mm) & $k$ & $A$ (mm$^{-3}$) & $B$ (mm$^{-5}$) \\
\tableline
Primary   & 2560          & -1.0 & 0.0 & 0.0\\
Secondary & 363.83041     & -0.882787413818 & 1.3641139$\times10^{-9}$ & 1.8691461$\times10^{-15}$\\
Tertiary  & 545.745477407 & -1.0 & 4.3908607$\times10^{-10}$ & -3.2605391$\times10^{-15}$\\
\tableline
\end{tabular}
\end{center}
\caption{
Ideal parameters for the equation of the three \boom mirrors
(see eqn.\ref{eqn:opt})}
\label{tab:mirrors}
\end{table}

\begin{figure}[ht]
\epsscale{.6}
\begin{center}
\rotatebox{90}{
\plotone{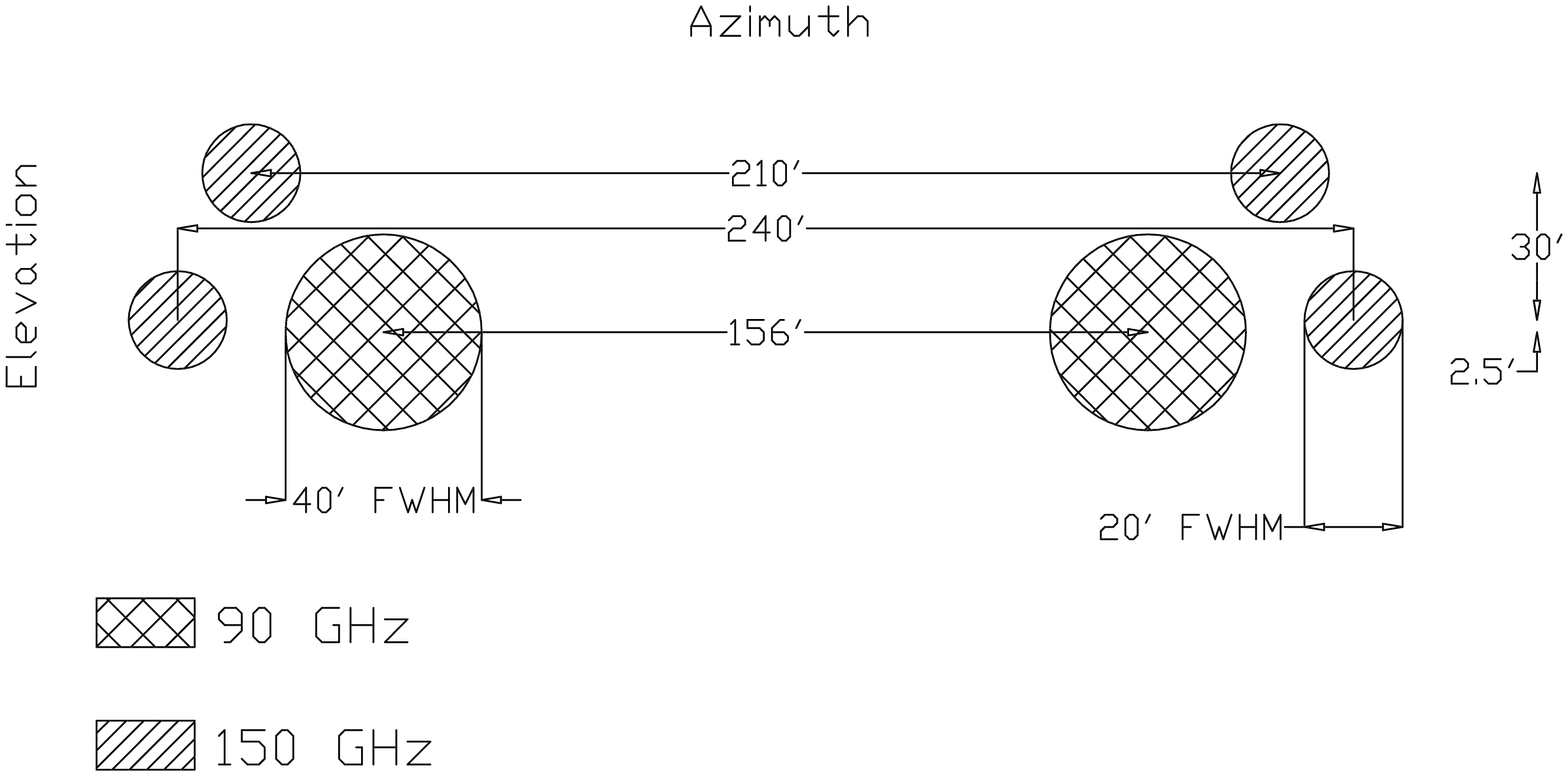}}
\caption{Location of the beams in the focal plane of the BNA photometer.
\label{fig:fp}}
\end{center}     
\end{figure}

The BNA focal plane contains single frequency
channels fed by conical or Winston horns. Although the
image quality from the optics is diffraction limited over a $2^{\circ}
\times 5^{\circ}$ field, all of the feed optics are placed inside
two circles $2^{\circ}$ in diameter, separated  by 
$3.5^{\circ}$ center to center.  The focal plane area outside these circles 
is vignetted by blocking filters at the entrance to the optics box
and on the 77~K shield and is unusable.  Due to the curvature of the
focal plane, the horns are placed at the positions of the beam centroids
determined by geometric ray tracing.  All of the feeds are oriented 
towards the center of the tertiary mirror.  The configuration of the
focal plane for the North American Flight is in Figure~\ref{fig:fp}.

The BNA frequency bands are centered at 90 and 150~GHz and are optimized
to maximize sensitivity to CMB fluctuations, identify dust emission
and reject radiation from atmospheric line emission. 
The low pass filters mounted on the 77~K and 2~K optical entrance windows 
provide good transmission across these bands while effectively blocking 
higher frequency radiation.  Two metal mesh low pass filters 
with cutoff frequencies
of 480~GHz are mounted on the 77~K stage.  The filters are 65~mm in
diameter and each one is directly behind one of the
50 $\mu$m polypropylene vacuum windows.  Two more low pass
filters with a cutoff frequency of 300~GHz are mounted on the 2~K
stage at the entrance to the cold optics box.  Although each filter
has a leak at twice the cutoff frequency, the combination eliminates
these leaks. Because these filters
are reflective at frequencies above the cutoff, they minimize the
radiation heat load on the cryogens.  While the metal mesh filters have
high reflectivity at high frequencies, they are not impenetrable, and above
a few THz any transmission
at the level of less than $10^{-3}$ is significant.  Therefore, we
place a dielectric absorber behind the 300~GHz filter on the 2~K stage
that has a cutoff frequency of 1650~GHz.  The absorber is 0.5~mm thick
alkali-halide filter coated with a 130~$\mu$m thick layer of 
black polyethylene.
This low pass filter stack has a transmission of $>80$\% at all frequencies
from 90 to 410~GHz, while attenuates by a factor
$5 \cdot 10^{-4}$ between 1650~GHz and 100~THz and by a factor $10^{-3}$
over 100~THz.
All of the other filtering is done in the focal plane elements.

\begin{figure}[th]
\epsscale{.5}
\begin{center}
\plotone{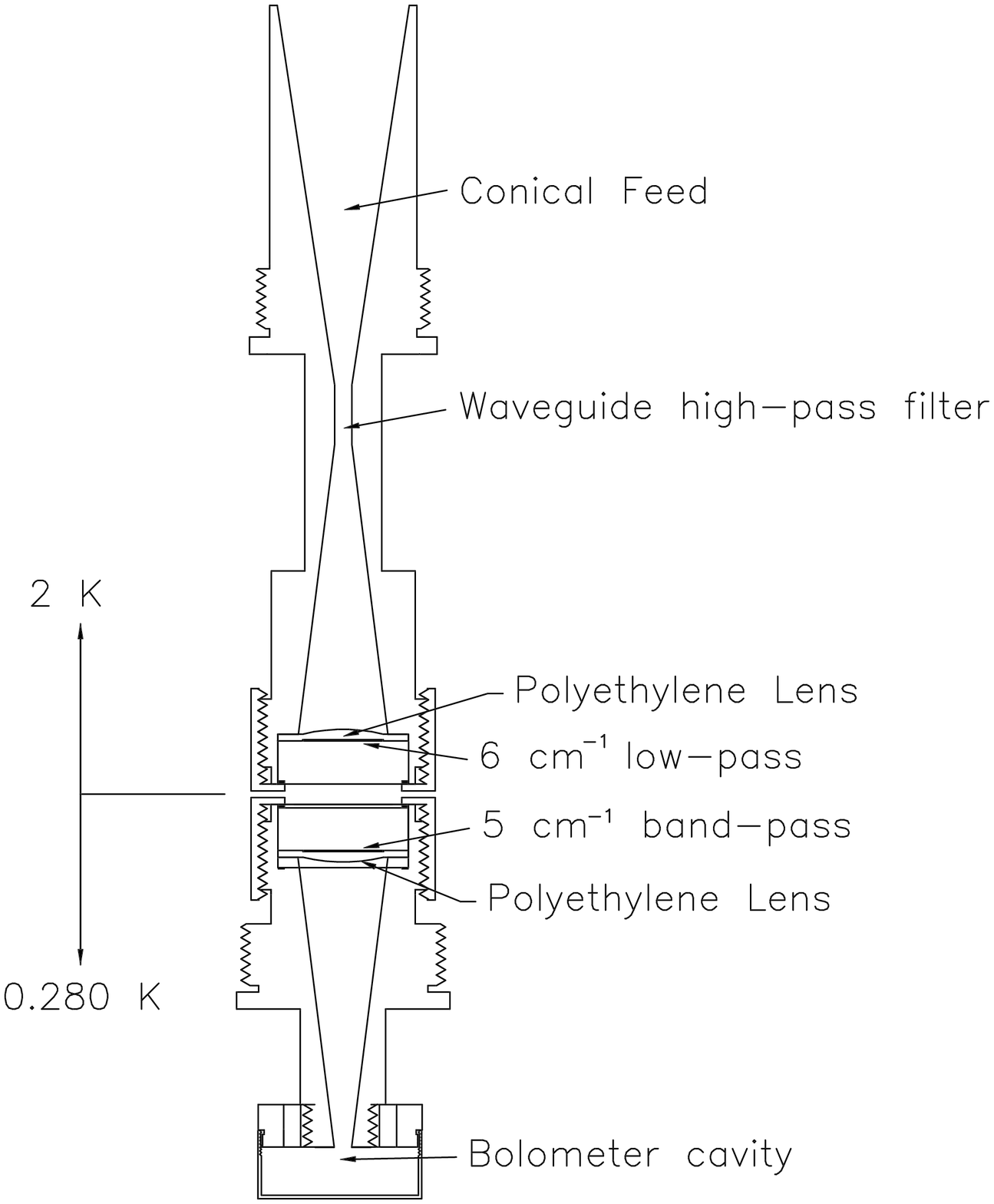}
\caption{
150~GHz feed horn. The upper part is cooled at less than 2~K and
the lower part to 0.280~K. See text for description. 
\label{fig:feed}}
\end{center}     
\end{figure}

We have produced
two different feeds designed to efficiently couple to the telescope and produce
beam sizes on the sky of $20^{\prime}$ and $40^{\prime}$.  
These sizes are larger than the diffraction limit: we traded resolution
for throughput to obtain good sensitivity to diffuse radiation even
in a short test flight.
The design of
the single frequency feed structure is shown in Figure~\ref{fig:feed} and is
similar to the feed design described in \cite{Church}. This design allows
us to illuminate correctly the band defining filters 
and create an effective Faraday cage surrounding the bolometric detectors.

Each feed
consists of a band defining filter stack mounted inside waveguide optics
that couple the radiation from the tertiary mirror into an integrating
bolometer cavity.  The feed is divided in two parts separated by a 0.5~mm
gap, one held at 2~K and the other at 300~mK.

The entrance horns are mounted in a horn positioning flange held at
2~K.  Each horn couples to a section of waveguide with length $2\lambda$
that acts as a high pass filter and rejects radiation with wavelength
$\lambda > 3.41 a$, where $a$ is the radius of the waveguide.
The $20^{\prime}$
horns have an entrance aperture of 19.7 mm at $f=3.3$ and a 
waveguide diameter of 2.54 mm.  
The throughput is
limited by a combination of the entrance aperture of the horns and the
size of the Lyot stop to be $A\Omega_{\rm total} = 0.1$~cm$^2$sr for the
$20^{\prime}$ horns. This value is over the diffraction limit throughput
$A\Omega_{\rm horn} \simeq\lambda^2=0.04$~cm$^2$sr at 
$\lambda \simeq 2$~mm (150~GHz). 
The $40^{\prime}$ horns have an entrance aperture of 33 mm 
and a waveguide diameter
of 5.1 mm.  These channels have approximately three times the throughput
of the $20^{\prime}$ channels with $A\Omega \simeq 0.3$~cm$^2$sr, while the
diffraction limit is 0.1~cm$^2$sr.

On the other side of the waveguide is an f/4 conical horn that expands
the waveguide to a diameter of 17.8 mm in the 20$^{\prime}$ 
channel and 27.9 mm in
the $40^{\prime}$ channel.  At the end of this cone, the
feed opens up to allow the mounting of a converging lens and an oversized 
metal mesh dichroic filter.  A band pass filter followed by another
identical converging
lens and conical horn are held at 300~mK across the small gap of $< 1$~mm.
This final horn feeds a small cavity containing a micromesh bolometer
which absorbs $> 90$\% of the radiation entering the cavity.  In the
geometrical limit, the lenses reimage to infinity a point at the vertex 
of the horns on which they are mounted.  In the diffraction limit, they
each produce a beam waist at the center of the length of waveguide
that contains the filters, thereby improving the coupling between the
expanding horn at 2~K and the concentrating horn at 300~mK.

\begin{figure}[ht]
\epsscale{.6}
\begin{center}            
\rotatebox{270}{
\plotone{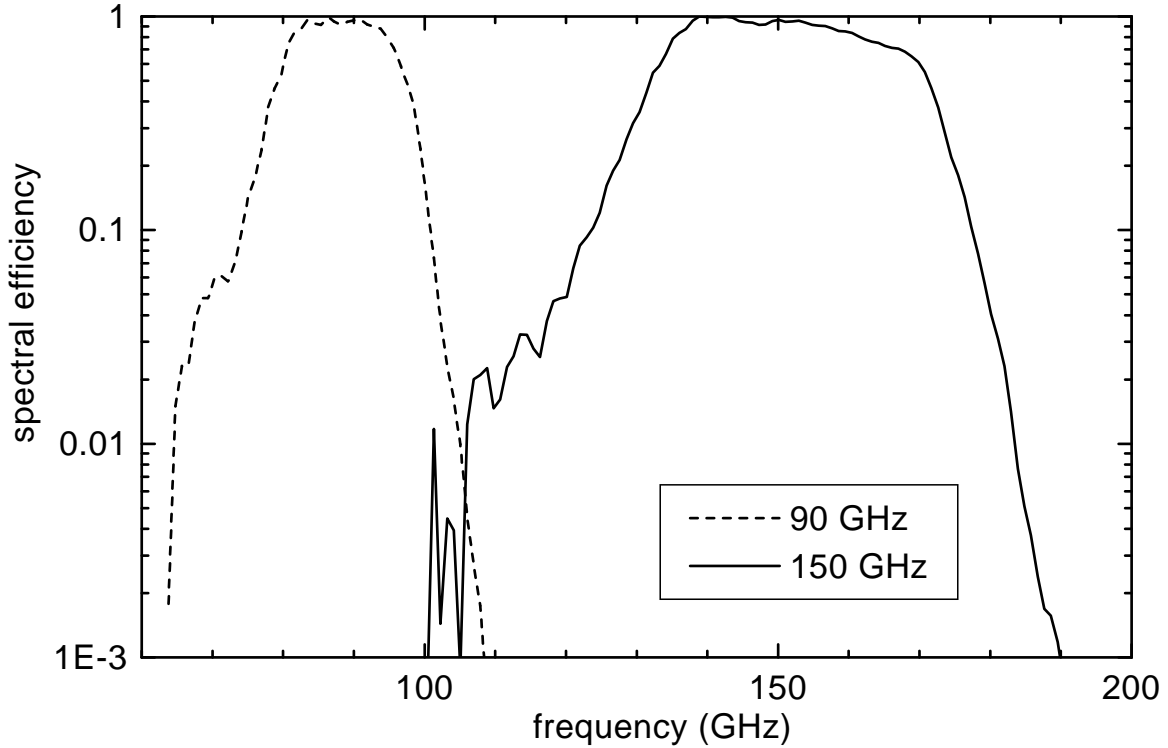}}
\caption{
Frequency response of a 90 and a 150~GHz channel as measured
with a Fourier Transform Spectrometer.
\label{fig:filter}} 
\end{center}
\end{figure}

The spectral bands are defined by the 300~mK band pass filters,
metal-mesh resonant grid filters with nominal central frequencies of 90~GHz
and 150~GHz and 30\% bandwidth.  These filters have high transmission
in band with sharp band edges, but their performance degrades quickly
as a function of off-axis angle.  The f/4 horns in the intermediate
section of the feed structure insure that the filters are illuminated
with small off-axis angles.
The band pass filters also have a leak at approximately twice the
central frequency.  The 2~K dichroic filters in the feed structure
eliminate these leaks and provide additional high frequency blocking
with cutoff frequencies of 172~GHz for the 90~GHz channel and 
217~GHz for the 150~GHz channel.  The frequency response
of a 90~GHz and a 150~GHz 
channel, measured with a
Fourier Transform Spectrometer, are shown in Figure~\ref{fig:filter}.  
The out-of-band transmission is $<-25$~dB
for $\nu<900$~GHz.

We have measured the optical efficiency of this structure equipped with
filters for both 90~GHz and 150~GHz operation.  The optical efficiency
measurements were made using black loads of eccosorb foam at 77~K and
300~K.  We measure the difference in optical power on a bolometer
by comparing the DC-coupled I-V curves of the detector under the
different loading conditions.
Pre-flight results are in Section \ref{sec:preflight}.

Tables \ref{tab:bands} and \ref{tab:loading} give the band centroids
and widths for the \boom photometers, along with the estimated
loadings from the CMB, the telescope, and the atmosphere.

\begin{table}[ht]
\begin{center}
\begin{tabular}{ccccc}
\tableline
\tableline
$\nu_{nominal}$&$\nu_{\rm peak}$ &$\nu_0^{\rm CMB}$ &
$\nu_0^{\rm R-J}$ & $\Delta \nu_{\rm FWHM}$ \\
(GHz) & (GHz) & (GHz) & (GHz) & (GHz) \\
\tableline
90 & 88.5 & 93.6 & 94.1 & 33 \\
150 & 136.5 & 155.7 & 157.9 & 54 \\
\tableline
\end{tabular}
\end{center}
\caption{
\boom North America Filter Bands. $\nu_{peak}$ is the peak
for a flat spectrum, $\nu_0^{\rm CMB}$ and $\nu_0^{\rm R-J}$
are respectively the peaks for a CMB and a Rayleigh-Jeans
spectra.}
\label{tab:bands}
\end{table}

\begin{table}[ht]
\begin{center}
\begin{tabular}{cccccc}
\tableline
\tableline
 & $\nu_{\rm peak}$ & $A\Omega$ & P$_{\rm atm}$ &
 P$_{\rm CMB}$ & P$_{\rm tel}$ \\
 & (GHz) & cm$^2$sr  & (pW) & (pW) & (pW) \\
\tableline
BOOM/NA & 88.5 & 0.3 & 0.004 & 0.10 &  0.42 \\
       & 136.5 & 0.1 & 0.28 & 0.17 & 1.6 \\
\tableline
\end{tabular}
\end{center}
\caption{\boomn/NA in flight expected loadings}
\label{tab:loading}
\end{table}

\subsection{Detectors}
\label{detect}

\boom uses bolometers to detect the fluctuations in
incoming radiation.  A bolometer consists of a broadband
absorber with heat capacity C, that has a weak thermal link G,
to a thermal bath at a temperature T$_{\rm b}$.  Incident radiation
produces a temperature rise in the absorber
that is read out with a current biased thermistor.
The sensitivity of a bolometer expressed in Noise Equivalent Power
(NEP) is given by:
\begin{equation}
{\rm NEP_{bolo}} = \gamma \sqrt{4k{\rm T_b}^2 {\rm G}}
\end{equation}
where $\gamma$ is a constant of order unity that depends weakly
on the sensitivity of the thermistor and $k$ is Boltzmann's constant.  
The dynamical equation
for the temperature of the thermistor can be expressed as:
\begin{equation}\label{eqn:eqbolo}
\frac{\rm dT_{bolo}}{\rm dt} = \frac{\rm P_{in}-G(T_{bolo}-T_b)}{\rm C}
\end{equation}
where $P_{in}$ is the incident power.
From this equation, we can see that bolometers have a finite
bandwidth limited by the time for the absorber to come to equilibrium
after a change in incident power, $\tau =$ C/G.
Previous balloon-borne bolometric receivers
have been limited in sensitivity or bandwidth by the properties of
the materials used for fabrication of the detectors.

Bolometers are also limited in sensitivity by external sources
of noise such as cosmic rays, microphonic disturbances, and 
Radio Frequencies Interference (RFI).
Of particular importance for \boom is the cosmic ray
rate on the Antarctic stratosphere, which is about an 
order of magnitude higher than at North American latitudes because
the magnetic field of the earth funnels charged particles to
the poles.  We have experience of cosmic ray hits from many 
bolometric receivers flown at balloon altitude at temperate 
latitudes.
The cosmic ray hit rate for the MAX experiment \cite{Alsop}
during a North American balloon flight was 0.14 Hz.  In order to
remove spurious signals from these cosmic ray interactions, cosmic
rays were identified and a length of data corresponding to 10
detector time constants was removed.  Because the MAX detector
time constants were all less than 30~ms, this resulted in a loss
of 5\% of the data which did not severely affect the sensitivity.
Almost half of the data would be contaminated by cosmic rays with
the same detectors operated from a Long Duration Balloon over the
Antarctic.

In the North American flight of \boom we tested for the first time
bolometers with a new architecture,
consisting of a micromesh absorber with an indium bump-bonded NTD 
germanium thermistor.  These bolometers can have lower heat capacity, 
lower thermal conductivities, lower cosmic ray cross section, and less 
sensitivity to microphonic heating than previous 300~mK bolometers.  
The micromesh absorbers and support structures are fabricated from a
thin film of silicon nitride using microlithography.  The absorber
is a circular grid with 60-400~$\mu$m grid spacing and 2-10\% filling
factor, metallized with 50~\AA \, of chromium and 200~\AA \, of gold.
It is designed to efficiently couple to millimeter wave radiation and
have low heat capacity and low cosmic ray cross section. Stiff mechanical
support consists in strands
of silicon nitride 1000~$\mu$m long with 3-5~$\mu$m$^2$ cross sectional
area connecting the absorber to a silicon frame, with a thermal 
conductivity of less than
$2 \times 10^{-11}$~W~K$^{-1}$.
The design and construction of the micromesh absorbers as well as their
optical and mechanical properties are described in
\cite{pdmoptics}.

\begin{figure}[ht]
\epsscale{.6}
\begin{center}
\plotone{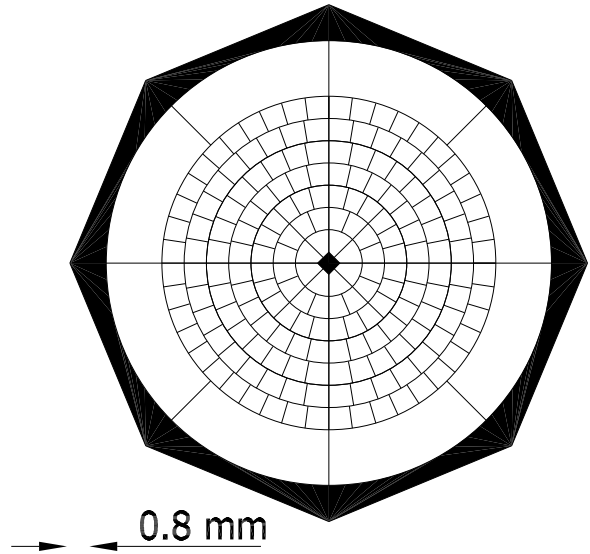}
\caption{
Schematic draw of a micromesh spider web bolometer absorber.
\label{fig:web}}
\end{center}
\end{figure}

The thermistor is a rectangular prism of Neutron Transmutation Doped
germanium 50~$\mu$m $\times$ 100~$\mu$m $\times$ 300~$\mu$m, a
factor of 10 times smaller volume than NTD themistors typically used
in composite bolometers.  Using these small thermistors decreases the
heat capacity of the bolometer by a factor of 5 and increases the
fundamental microphonic frequency by a factor of 10.
To allow the thermistor to be indium bump-bonded to the micromesh, one of
the long faces of the NTD material is metallized with two pads 50~$\mu$m
wide at either end.  Each pad is first implanted with boron and then
sputtered with 200~\AA \, of palladium and 4000~\AA \, of gold.
Electrical leads, gold pads, and indium bumps are patterned on the
micromesh while the silicon nitride still has a solid backing of silicon.
Gold pads and indium bumps are deposited at the center of the micromesh on
a 300~$\mu$m $\times$ 300~$\mu$m solid square of silicon nitride.
For electrical leads, 200~\AA \, of gold is deposited on some of the
support legs of the silicon nitride connecting the pads for the
thermistor with large gold pads on the silicon frame.
The thermistor is pressed onto the indium bumps and finally
the silicon is etched from the back of the micromesh.  The front of
the silicon nitride is coated in wax to protect the chip and the
metallization during etching.  This technique minimizes the amount
of material used for reading out the thermistors and therefore
minimizes the heat capacity of the device.  The indium bump-bonds
have survived repeated thermal cycling.  A draw of a micromesh
bolometer is shown in Figure~\ref{fig:web}. 
The performance of bolometers with indium bump-bonded thermistors 
is described in \cite{bock}.

\subsection{Readout Electronics}
\label{readout}

Most bolometric detectors in use today employ a high-impedence
semiconductor thermistor biased with a constant current.
JFET preamplifiers have been used to provide a combination of low
voltage and current noise well matched to typical bolometer
impedences \cite{halpern}, but exhibit excess voltage noise at
frequencies typically below a few Hz and have limited the achievable
bandwidth of these DC biased bolometers.  At lower frequencies, drifts
in the bias current, drifts in the temperature of the heat sink, and
amplifier gain fluctuations have been expected to limit the
ultimate stability of single bolometer systems.

AC bridge circuits have been successfully used in many experiments 
to read out pairs of 
bolometers with stability to 30 mHz~\cite{Wilbanks,MAX4b}.  
In this scheme a pair of detectors is biased with an alternating 
current so that resistance fluctuations are transfomed into changes 
in the AC bias amplitude across each detector.  
These signals are differenced in a bridge, amplified
and demodulated.  The AC signal modulation eliminates the effects
of 1/f noise in the preamplifiers since the resulting signal spectrum
is centered about the carrier frequency. The effects of drifts in the
bias amplitude, amplifier gain, and heat sink temperature are
greatly reduced if the two detectors in the bridge are well matched.
The optical responsivity of each of the detectors in the bridge is equivalent
to that of a detector biased with the same rms DC power and is
constant in time as long as the average power on the detector is
constant over the course of one detector thermal time constant.

\begin{figure}[ht]
\epsscale{.9}
\begin{center}
\plotone{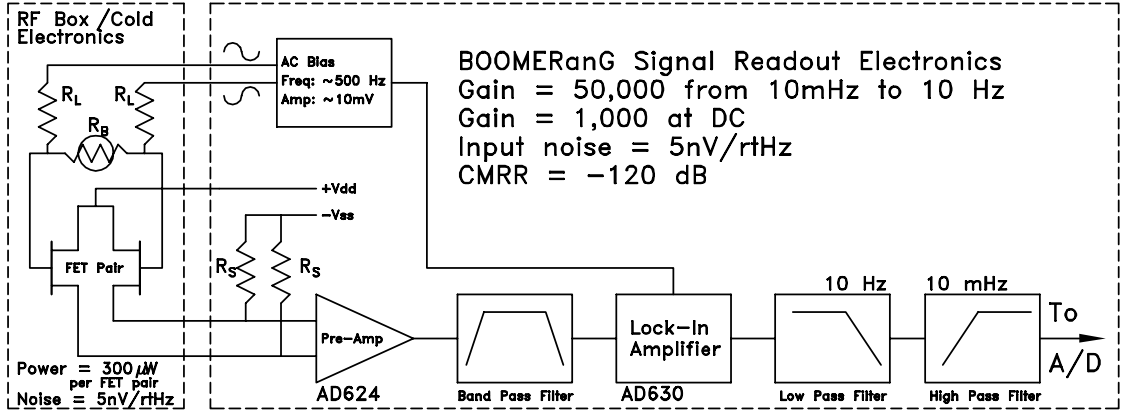}
\caption{
Block diagram of the bolometers readout electronics
(one channel shown). The bolometer is AC biased with a differential,
low-pass filtered square wave at $\sim$500~Hz.
The AC voltage across the bolometer is modulated by the resistance
variations induced by changes in the microwave power absorbed.
A matched pair of low noise J-FETs inside the cryostat reduce the signal
impedance from $\sim 10 {\rm M}\Omega$ down to $\sim 1 {\rm k}\Omega$;
the signal
is then amplified by a differential preamp (AD624), band-pass filtered to
remove noise outside the signal bandwidth, and synchronously demodulated
by a phase sensitive detector (AD630). The output of the AD630 is
proportional to the instantaneous resistance of the bolometer.
Signal components below 10~mHz are attenuated to get rid
of 1/f noise and drifts using a single pole high pass filter. High
frequencies (above 10~Hz, i.e. above the cutoff frequency of the
bolometer) are also removed by means of a 4-th order low-pass filter. The
resulting signal is analog to digital converted with 16~bits
resolution, at a sampling frequency of 62.5~Hz.
\label{fig:readout}}      
\end{center}
\end{figure}

\boom employs an AC stabilized {\it total power} readout system
for individual bolometers, mounted on a temperature regulated stage 
\cite{hristov1}. The circuit is summarized in 
Figure~\ref{fig:readout}.

This system contains a cold J-FETs input stage 
(based on Infrared Laboratries TIA) and contributes less than 
10~nVrms/$\sqrt{\rm Hz}$ noise at all frequencies within the bolometer signal
bandwidth down to 20~mHz.  The warm readout circuit has a gain stability of
$< 10$ ppm/$^\circ$C. We remove the large offset due to the background
power on the detector with a final stage high pass filter with a cutoff
frequency of 16~mHz.  With this circuit, we have measured the noise spectrum
of a low background micromesh bolometer with 
NEP~$=1.2 \times 10^{-17}$~W$/\sqrt{\rm Hz}$, biased 
for maximum responsivity to be flat 
down to a frequency of 20~mHz.

\subsection{RF Filtering}

There are many sources of Radio Frequency Interference (RFI) on the 
balloon that could couple to bolometers.
Microwave transmitters (400~MHz to 1.5~GHz, few W)
that send the data stream to the ground
and high current wires that drive the motors of the Attitude Control
System (20~kHz PWM, several Amps) are situated within a few meters 
of the cryostat.
The \boom wiring and focal plane are designed to prevent RFI
from contributing to the noise of the bolometers. 

The bolometers are contained inside a 2~K Faraday cage inside
the cryostat.  RFI can enter the cryostat through the optical
entrance window and propagate into the 2~K optics box.  However,
the exit aperture of the optics box is RF sealed by the horn positioning
plate.  This plate contains feed horns with small waveguide apertures
for radiation from the sky to pass through to the detectors.
The largest waveguide feedthrough in this plate is 5.1 mm in diameter
which corresponds to a waveguide cutoff of $\sim35$~GHz.  Lower frequency
RFI is reflected by this surface.  Readout wires entering the
bolometer Faraday cage can also propagate RF signals as coaxial cables.
We run all of the bolometer wires through cast eccosorb filters mounted
to the wall of the Faraday cage to attenuate these signals.  The
filters are 30~cm long and have a measured attenuation of $< -20$~dB
at frequencies from 20~MHz to a few~GHz.

The readout electronics are also sensitive to RFI.  We enclose all
of the cryostat electronics in an RF tight box that forms an extension
of the outer shell of the cryostat.  The signals from the detectors
pass through flexible KF-40 hose that is RF sealed to the hermetic connector
flange on the cryostat and to the wall of the electronics box.
The amplified signals exit the electronics box through {\it Spectrum}
RF filters mounted on the wall of the box. 

\subsection{Cryogenics}
\label{cryo}

\begin{figure}[ht]
\epsscale{.6}
\begin{center} 
\rotatebox{90}{
\plotone{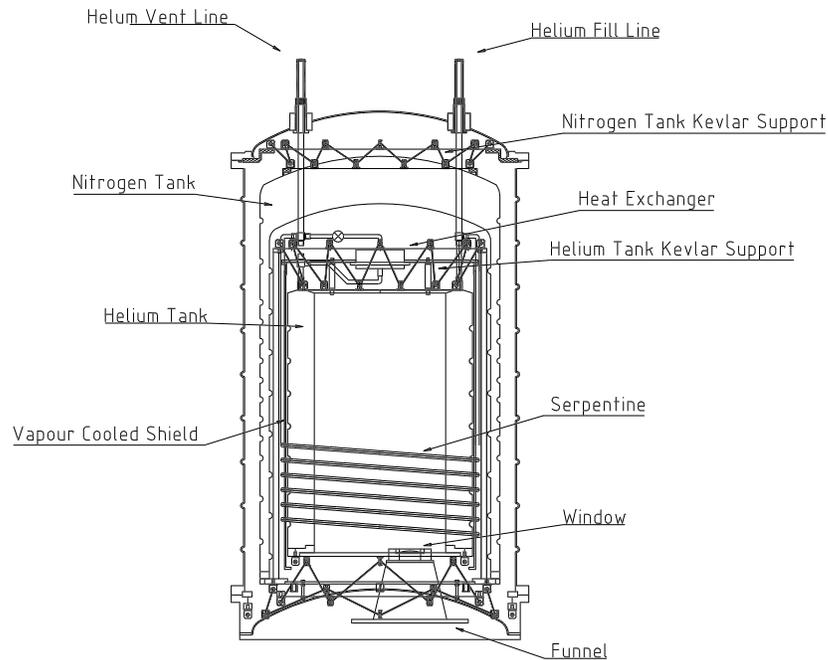}}
\caption{
Main cryostat. The evaporating $^4$He
gas flow through the serpentine and
the heat exchanger to cool a copper shield that minimize the
heat input on the main Helium bath. The tanks are supported
by kevlar cord. Nitrogen lines are not shown.
\label{fig:cryostat}}
\end{center}
\end{figure}

A ``heavy duty'' $^3$He fridge and a large $^4$He cryostat have been 
developed specifically for the \boom experiment. A cutaway view
of the cryostat is shown in Figure~\ref{fig:cryostat}.  

The main $^4$He 
cryostat has to be large enough to contain refocusing optics and a
wide focal plane with several multiband photometers. The design and
performance of the cryostat and refrigerator are described in detail
elsewhere \cite{fridge,cryostat}.  The total volume
occupied by the cryogenic section of the receiver is 69~liters.
The design hold time is about 20 days; the helium tank volume is
60~liters, the nitrogen tank volume is 65~liters. Conduction thermal input 
is reduced by suspending both the tanks with Kevlar ropes (1.6~mm 
diameter). The vibration frequencies of these structures are all
above 20~Hz, and the amplitude of the vibrations excited during
the flight is expected to be very small.
Radiation thermal input on the nitrogen tank is reduced by 
means of 30 layers of aluminized mylar for superinsulation. 
The total thermal input on the nitrogen bath is $6.6\pm 0.4$~W.

The radiative thermal load on the L$^4$He is minimized by the use of a
vapor cooled shield.  As the liquid helium evaporates, the cold gas
flows through a
spiral tube that is soldered to the outside of a copper shield which
surrounds the Helium tank and through a copper heat exchanger that is attached
to the top of the vapor cooled shield before emerging from the cryostat.
The temperature of the vapor cooled shield depends on the gas flow rate
from the Helium tank. During normal operation, the shield remains at
a temperature of 15-17~K.
The total thermal input on the helium bath is linked to the radiative input 
through the cryostat window and in flight conditions is about 75~mW.

The cryostat has two circular windows 66~mm in diameter, made 
with 50~$\mu$m polypropylene supported by an aluminum frame.

In the BNA flight we used Indium seals able to work at the very low
temperature experienced during the night flight ($-50^\circ C$). In the 
day-time LDB flight it was possible to use rubber (Buna-N) seals being 
the temperature of the dewar always $> -30~^\circ$C.

\begin{figure}[ht]
\epsscale{.5}
\begin{center}
\plotone{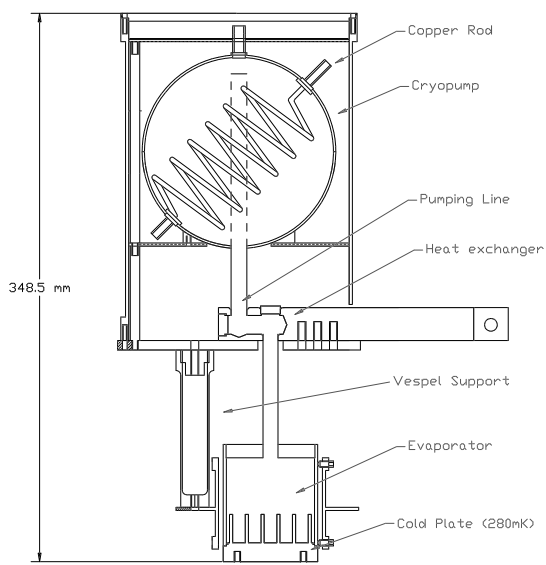}
\caption{
\boom $^3$He fridge. See description in the text. A mechanical thermal
switch, not reported in the drawing, connects the cryopump to the heat
exchanger flange.
\label{fig:fridge}} 
\end{center}
\end{figure}

The bolometers are cooled by a self contained $^3$He fridge
shown in Figure~\ref{fig:fridge}.
The fridge operates with a charcoal sorption pump and can be
recycled electronically.  During a cycle, hot $^3$He gas is expelled
from the pump by heating it to $\simeq 40$~K with a resistance heater.
This gas passes through a thin-walled stainless steel tube and
condenses on a high purity copper condensation surface at $\simeq
2$~K.  The $^3$He liquid then drops through another stainless
steel tube into a 125~cm$^3$ evaporator pot.  Finally, when all
of the gas has collected in the evaporator, the heater power is
removed from the pump and it is connected to the condensation
point with a mechanical heat switch. The pump temperature
drops to $\sim 3$~K and pumps on the liquid $^3$He in the
evaporator, reducing its temperature to $< 300$~mK.

The pump is made of two stainless steel hemispheres crossed by a
copper rod that extends outside, to provide thermal attachment
points to both the heater and the thermal switch.  
The pump is connected to the copper condensation plate ($\sim 2$~K)
through a 1~cm diameter stainless steel tube.  There is a 1~cm
diameter hole that horizontally passes through the condensation plate, 
with a 1~cm path to provide more surface area to transfer
heat from the $^3$He gas to the copper.  The evaporator consists of
a stainless steel cylinder with a 6~cm diameter copper endcap with
a ring of 6~mm diameter bolt holes for thermal attachment.
The charge is 34~liters STP at 40~bars.
The heat load on the $^3$He stage is about $20~\mu$W.
About half of this is due to thermal conduction from the condensation
point to the evaporator through the 0.010'' diameter stainless steel
pump tube.  The remaining 10~$\mu$W is from the mechanical supports
for the focal plane.  The focal plane weighs 2.5~Kg and must be
rigidly held with a mounting structure strong enough to withstand
10~G acceleration in any direction while providing minimal additional
thermal input to the $^3$He fridge.  We use four thin-walled vespel
tubes to satisfy these requirements.  Vespel has an extremely high
strength to thermal conductivity ratio at temperatures from 0.3 to 2~K.
The tubes are 1'' in diameter, 0.030'' thick and 3'' long.  The yield
strength of a tube can be calculated from standard formulae.
The criterion we use for strength is that the maximum deflection be
less than 1\% of the elastic limit of the material.  Additional heat
load from electrical leads is minimized by the use of 0.005'' diameter
manganin wires that run up the length of the vespel tubes between
the cold JFET amplifier box and the detectors.  The wires are firmly
attached to fixed surfaces along their entire length with teflon tape
to eliminate vibrations which can make the detectors microphonically
sensitive.  The length of the wires is minimized to reduce their
contribution to the input capacitance at the JFETs.  The RC cutoff
frequency for a 5~M$\Omega$ detector impedence is measured to be
$>2$~kHz, for an input capacitance of $<40$~pF.

During observations, the temperature of the focal plane can be
maintained constant with a high precision temperature regulation 
circuit \cite{hristov2}. In fact,
temperature fluctuations of the 300~mK stage can contribute to excess
bolometer noise.  The temperature of the bolometer is determined by:
\begin{equation}
{\rm T_{bolo}} = {\rm T_0} + \frac{\rm P_{opt}}{\int_{\rm T_0}^
{\rm T_{bolo}} \kappa_0 {\rm T}^{\alpha} {\rm dT}}
\end{equation}
where ${\rm P_{opt}}$ is the optical power on the detector and
$\int_{\rm T_0}^{\rm T_{bolo}} \kappa_0 {\rm T}^{\alpha} {\rm dT}$,
indicated as G$_{\rm eff}$ (W~K$^{-1}$) hereafter, 
is the effective thermal conductivity per unit
of temperature between the thermistor and the bath at ${\rm T_0}$.
Therefore the change in bolometer temperature for a change in base
plate temperature is given by:
\begin{equation}
\Delta {\rm T_{bolo}} = \beta \Delta {\rm T_0} 
\end{equation}
where $\beta$ is a constant of order unity.  The bolometer NEP is related
to these temperature fluctuations by:
\begin{equation}
{\rm NEP_T} = {\rm G_{eff}}\Delta{\rm T_{bolo}}
\end{equation}
where ${\rm NEP_T}$ is in W/$\sqrt{\rm Hz}$
and $\Delta {\rm T_{bolo}}$
(K/$\sqrt{\rm Hz}$) is the spectrum of temperature fluctuations.  Therefore,
the condition for temperature stability of the cold stage is NEP$_{\rm T} <
{\rm NEP_{bolo}}$.  For the \boom bolometers with G$_{\rm eff} = 8
\times 10^{-11}$~W~K$^{-1}$ and ${\rm NEP_{bolo}} \simeq 2 \times
10^{-17}$~W/$\sqrt{\rm Hz}$
the requirement for the stability of the cold stage is:
\begin{equation}
\Delta {\rm T_{bolo}} < 250 {\rm nK}/\sqrt{\rm Hz}
\end{equation}

Two NTD thermistors are mounted on the 300~mK stage and read out
with DC coupled bridge circuits.  
One channel is used as the control sensor and the other is a monitor sensor.  
We measure the spectrum of fluctuations
from the monitor channel while the temperature regulation circuit
is running to be flat down to $<30$~mHz with an amplitude of
40~$\mu$Vrms/$\sqrt{\rm Hz}$ corresponding to a temperature stability
of 120~nK/$\sqrt{\rm Hz}$.
This is better than is
needed to insure that the \boom bolometers do not have a significant
noise contribution from temperature fluctuations.

\subsection{Pressure control systems}

The pressure on the main helium bath is maintained around 10~mbar 
in lab and in flight to 
keep the bath temperature below 2~K, while the  
pressure on the nitrogen bath is kept around 1~atm to
prevent the formation of fluffy solid. 
We used two pressurization systems. The helium
pressurization system allows us to pump on the bath at ground
and to open to atmosphere pressure at float by means of a motorized
high vacuum seal valve.
The nitrogen system controls the pressure using an absolute 
sensor and three electrovalves.
Both systems have relief valves to allow safety and functionality 
even in the case of electronic failures.

\subsection{Internal calibrator}
\label{internalcal}

\begin{figure}[ht]
\epsscale{.6}
\begin{center}  
\plotone{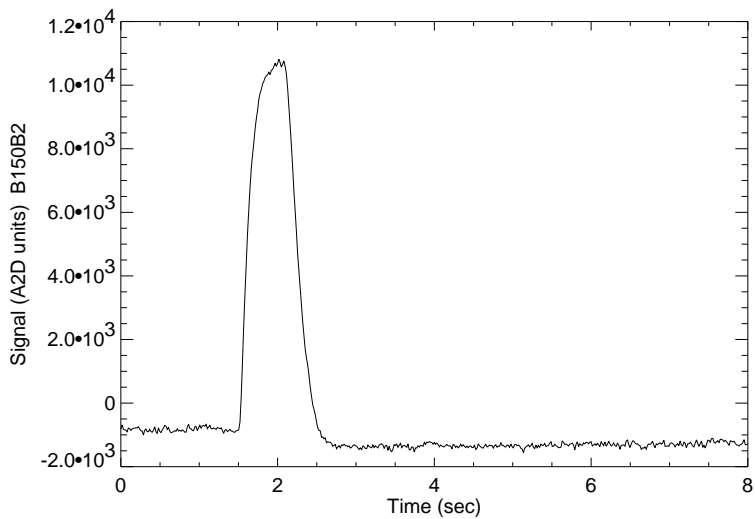}
\caption{
Bolometer signal during a pulse from the internal calibrator.
A/D units are used, corresponding to a full scale of
$\pm 10$~Volts in 16~bits.
\label{fig:callamp}}
\end{center}
\end{figure}

An internal calibrator in the reimaging optics box allows us to 
monitor the detector sensitivity during flight.  The calibrator 
consists of a high background bolometer with an NTD~4~Ge thermistor 
attached to the center of a 5.6~mm diameter Silicon Nitride absorber,
electrically connected with 0.001'' diameter copper leads. The calibrator is 
mounted on the 2~K stage directly behind a 1~cm hole in the center 
of the tertiary mirror, where all of the beams from the concentrating 
horns are coincident.  The calibrator can be heated with a current 
pulse to give a temperature rise of several Kelvin with a 
recovery time of a few milliseconds.  The corresponding power on the 
detectors is:
\begin{equation}
{\rm P_{cal}} \simeq 2 k \Delta {\rm T_{cal}} \frac{\nu^2}{c^2}
\Delta \nu \cdot \Omega \epsilon f
\end{equation}
where $\epsilon$ is the calibrator efficiency and $f$ is the 
effective Lyot stop fraction area filled by the calibrator.
This corresponds to an equivalent signal on the sky of
\begin{equation}
\Delta {\rm T_{CMB}} = \epsilon \Delta {\rm T_{cal}} f
\end{equation}
Response to the internal calibrator
during flight for one of the detectors is shown in Figure~\ref{fig:callamp}.
The equivalent signal on the sky is $\sim 175~$mK. 

\section{Attitude Control System}
\label{acs}

The Attitude Control System (ACS) must be 
able to point in a selected sky direction, and track it
or scan over it with a reasonable speed. The specifications are 
1~arcmin~rms for pointing stability, with a reconstruction capability 
better than $\sim 0.5$~arcmin maximum. Our main modulation
is obtained scanning in azimuth, with a saw-tooth scan, with
an amplitude of 40~deg~(p-p) and a scan rate of about
2~deg~s$^{-1}$.  We have developed an ACS for \boom with these 
capabilities based on the ACS systems designed and built for the 
ARGO and MAX-5 experiments \cite{argoexp,MAX5}.
The \boom ACS is based on a pivot which decouples the payload
from the flight chain and controls the azimuth, plus one linear 
actuator controlling the elevation of the inner frame of the payload. 
The pivot has two flywheels, moved by powerful torque motors 
with tachometers. On the inner frame, which is steerable in elevation
with respect to the gondola frame, are mounted both
the telescope and the cryogenic receiver. The observable elevation range
is between 33 and 65~deg. The sensors are different
for night (North America) and day (Antarctic) flights. For night
flights we have a magnetometer and an elevation encoder; additional 
information on the average attitude is obtained by means of a sensitive 
tilt sensor. A CCD star camera is used outside the feedback loop for 
absolute attitude reconstruction. 
A CPU handles commands and observation
sequencing; the same CPU digitizes sensor data, and controls the 
current of the three torque motors.  

\subsection{CCD Star Camera}
\label{ccd}

We used a video CCD camera ({\it Cohu 4910})
with a large aperture lens ({\it Fujinon CF50L})
as a star sensor.
The focal length is 50~mm, the numerical aperture is $f/0.7$.
The CCD format is $1/2^{\prime\prime}$ and the video signal is
{\it RS-170} at 60~$Hz$.  The image area in the CCD is
$6.4 {\rm mm} \times 4.8 {\rm mm}$, with $768 \times 494$ pixels. 
The resulting field angle is $ 7.30^\circ \times 5.50^\circ$. 
The optics are baffled to protect the lens input against stray rays
from reflecting surfaces in the payload. The baffle is a corrugated 
cylindrical structure painted black on the internal surface and
dimensioned for complete protection of the lens input.
We removed the CCD protection glass in order to accommodate the lens 
output surface close enough to the CCD chip surface. 
Boresight between the star camera and the millimeter wave telescope
is trimmed at ground, and cannot be changed in flight.
The CCD is mounted on the gondola through fiberglass supports and is
surrounded by a protective and thermally insulating foam box. 
A few resistors, dissipating a total of 
$\sim 10 \ $W, are used to keep the system warm. 

CCD images are processed on board in real time.
We used a {\it Matrox Image-LC} board as a frame grabber and signal 
processing unit (DSP). The board accepts an analog (video) input and 
returns a VGA output image.
The star sensor computer is connected to the other flight computers 
through serial ports.
These electronics were assembled inside a pressure vessel in order 
to let the system operating at standard pressure during the flight. 
This is required for both the correct operation of the hard disk and the 
thermalization of the components.
A thermostat and two fans inside the vessel
control the internal air temperature. The total heat
dissipation of the system is $\sim 50$~W. When the fans are operating 
the heat dissipating devices are thermally connected to the outer  
shell of the vessel, radiating away the heat and effectively cooling
the system. If the air gets too cold, the fans stop, effectively insulating 
the system. In this way we can maintain
the operation temperature close to 20$^o$C during the flight, despite
of the low temperature ($\sim$~230~K) and pressure  ($\sim$~3~mbar).
We used the {\it Matrox Imaging Library (MIL)} package for the in flight
star recognition 
software in an optimized {\it C} code. The image analysis algorithm 
uses {\it blob analysis} to select the two brightest stars in the
field. The system is able to compute and transmit 
the pixel and brightness information five times per second.  

Before the flight we tested the performances of the CCD and of the 
optics in a vacuum chamber at room temperature and at $T \simeq 240$~K. 
We found a small increase of the sensitivity of the CCD at lower temperature, 
while there is no detectable defocusing of the image. 

The sensitivity of the camera was checked observing a star field 
from the Campo Imperatore Astronomical Observatory (at an altitude 2200~m 
above the sea level). It was possible to identify 
sources of magnitude \ $m \leq 6.5$, not far from the sensitivity we 
obtained during the flight.

When tested with suitable sources (i.e. stars with m~$\leq$~4 and
negligible seeing, as at float), the camera produces very accurate
positions of the target. Taking advantage of averaging insite in 
centroid determination, the accuracy of the coordinates
of the selected target is close to 1/10 of a pixel ($\sim$~3~arcsec).

The alignment of the CCD axis with the microwave beam was obtained by
observing a strong chopped thermal source placed at a distance of 
$\sim 120$~m and correcting for the parallax due to the offset
between the mm-wave telescope and the camera (540~mm in the 
meridian axis, 640~mm in the horizontal axis).
We also tested the tilt of the CCD field respect to the horizontal axis
performing pure azimuthal scans on the same source. We reduced 
the tilt below 2-3 pixels along a $6^\circ$ azimuthal scan. 

Since not all frames have two stars present, the pointing
was recontructed between good frames by integrating the
data from the gyroscopes (sampled at 10~Hz).  This led to a precision
of less than 1~arcmin rms in the pointing solution.

\begin{figure}[ht]
\epsscale{.6}   
\begin{center}  
\plotone{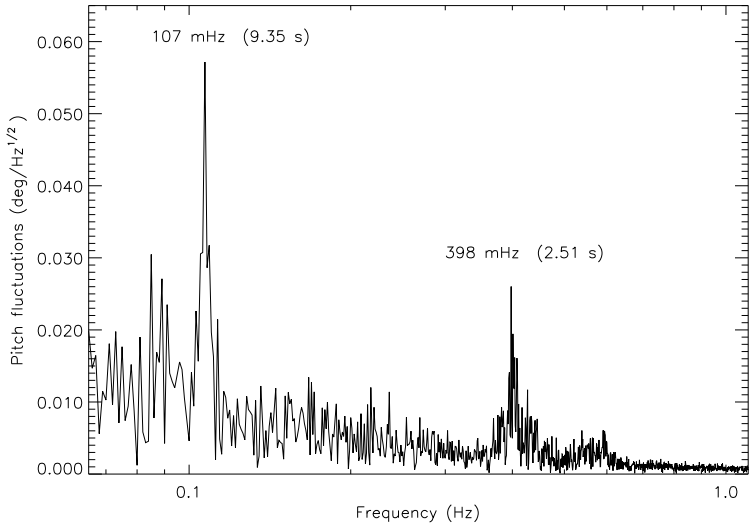}
\caption{
Power spectrum of the pitch of the payload during scans.
Small pendulations are at 107 and 398~mHz
corresponding to periods of 9.35 seconds (full flight chain) and
2.51 seconds (payload).
\label{fig:pend}}
\end{center}
\end{figure}

Pendulations of the gondola was monitored by roll and pitch
gyroscopes. The typical power spectrum of the pitch fluctuations 
is reported in figure~\ref{fig:pend}.

\section{Scan strategy}
\label{scans}

The \boom scan strategies are designed to obtain high signal to noise
measurements of CMB anisotropies covering as wide a range of angular
scales as possible, limited only by the functional bandwidth of the
detectors. The maximum scan speed is limited by the mechanics of
the Attitude Control System (ACS) and by the thermal time constant of
the bolometers.  The minimum scan speed is set by the
stability of the detectors, readout electronics, and sources of
local emission.  The ratio of these maximum and minimum signal
frequencies for the \boom detectors is $\approx 30$ allowing
us to cover a range in angular scale, corresponding to 
multipoles $10 < \ell < 900$.

The range of spatial frequencies, or multipole number $\ell$, that
we sample depends primarily on the beam size.  Because we use
bolometric detectors, we are free to select any beam size at
a given wavelength equal to or larger than the diffraction limit,
$\theta_{\rm diff} = 1.22\lambda/{\rm D}$ where D is the diameter
of the illumination pattern on the primary that is fixed for all
channels by the Lyot stop to be 85~cm.  The throughput for such a
system is a function of the beam size:
\begin{equation}
A\Omega(\lambda, \theta) = \lambda^2 \left(\frac{\theta}{\theta_{\rm diff}}
\right)^2
\end{equation}
The optical loading on the detector and the optimal thermal conductance, G,
are proportional to the throughput, $A\Omega$.  The absorbing area
of the detectors also must be proportional to $A\Omega$, however the
heat capacity, C,  of the micromesh bolometers is dominated by the thermistor,
so it is independent of throughput.  Therefore, the time constant of
optimized bolometers decreases as the beam size increases, $\tau \propto 
1/(A\Omega)$.  In addition, the signal from the CMB is proportional
to the throughput but the detector noise increases like the square root,
so that the sensitivity to CMB fluctuations improves like $\sqrt{A\Omega}$.
Therefore, overmoded bolometers are faster and more sensitive than
diffraction limited channels.  
We can make use of this to scan faster and cover a larger area
of sky with overmoded channels with high sensitivity and scan slower and
obtain higher angular resolution with diffraction limited channels.
We have optimized each \boom focal plane for angular resolution
and sensitivity.

\begin{figure}[ht]
\epsscale{.34}
\begin{center}
\rotatebox{90}{
\plotone{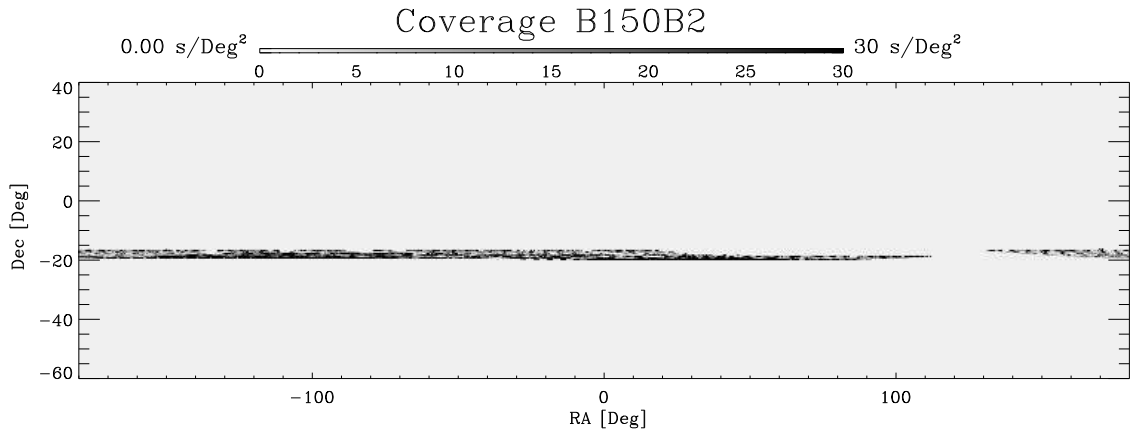}}
\caption{
Sky coverage and integration time distribution during the
flight of 1997, August 30.
\label{fig:coverage}} 
\end{center}
\end{figure}

The \boom North American flight, in 1997 August, covered a
region of sky approximately 700 square degrees (figure~\ref{fig:coverage}).  
This region was selected to have a low column density of
dust as estimated from the IRAS 100~$\mu$m measurements and provided even sky
coverage with an appropriate integration time per pixel based on
our detector sensitivity estimates.  The focal plane
shown in Figure~\ref{fig:fp} contained two 40$^\prime$ pixels 
at 90~GHz and four 20$^\prime$ pixels at 150~GHz. 
The gondola was scanned continuously in a triangle wave
with peak to peak amplitude of 40 degrees and a maximum scan velocity of 
2~deg~s$^{-1}$ corresponding to $\sim$2 bolometer time constants per beam.

\section{Pre-Flight Calibration}
\label{sec:preflight}

\begin{figure}[ht]
\epsscale{.6}
\begin{center}
\plotone{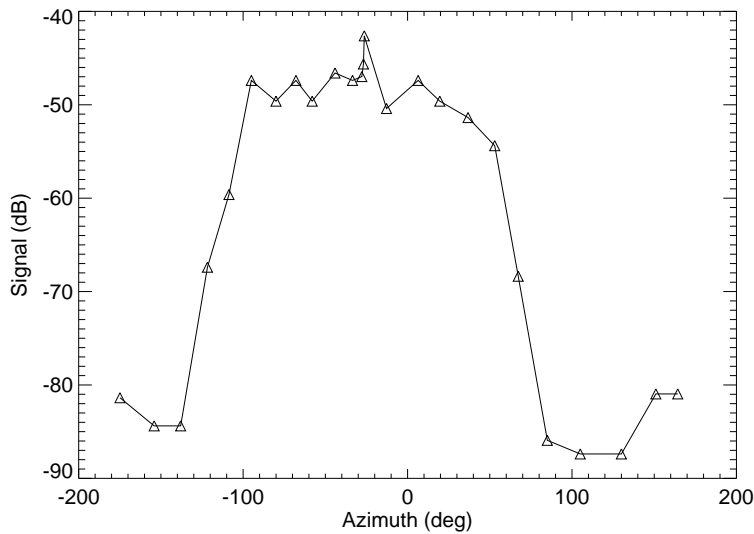}
\caption{
Sidelobes measurements at ground. The source is positioned at an
elevation of 39~deg, while the telescope is pointed at 49~deg
during this scan.
The sharp cutoff at large angles is due to the effect of the
large sun shields.
\label{fig:slobes}} 
\end{center}
\end{figure}

The pre-flight calibration procedure is performed before declaring 
flight-readiness, in order to have a precise forecast of the flight 
performance of the instrument.
For our telescope and receiver the most important tests are:

\begin{enumerate}
\item Bolometers load curves under different radiative loads. This is 
performed filling the photometer beam with blackbody loads at room 
temperature and at 77~K, and inserting in the beam a cold attenuator,
with 1\% transmission, to simulate the in-flight radiative background. 
Results for throughput and optical efficiency are in Table \ref{tab:opt}.

\begin{table}[ht]
\begin{center}
\begin{tabular}{lcc} \tableline
\tableline
Channel & throughput (cm$^2$ sr) &  optical efficiency \\ \tableline
NA-B150A1      &   0.11    &        8\% \\
NA-B150A2      &   0.08    &       19\% \\
NA-B150B1      &   0.11    &        8\% \\
NA-B150B2      &   0.08    &       20\% \\
NA-B90A        &   0.27    &        7\% \\
NA-B90B        &   0.27    &        9\% \\ \tableline
\end{tabular}
\end{center}
\caption{Pre-flight throughput and optical efficiency
calibration for the six detectors}
\label{tab:opt}
\end{table}

\item Voltage noise measurements of the system were performed to estimate 
the sensitivity and to check for 1/f noise in the detectors. Results in 
Table \ref{tab:noise}. Also a preliminary measure of the detectors
time constants $\tau$ was done to have an indication of 
performance. This measurement has to be repeated in flight in order to
have the correct values, strongly dependent on the background radiation
(see section \ref{time_c})

\begin{table}[ht]
\begin{center}
\begin{tabular}{lccccc} \tableline
\tableline
Channel & Noise & 1/f knee & Responsivity & NEP & NET$_{CMB}$     \\
   & $nV/\sqrt{Hz}$ & Hz  & V/W & $W/\sqrt{Hz}$ & $\mu K\sqrt{s}$\\
\tableline
NA-B150A1 & 12 & 0.8  & $4.5\cdot10^8$ & $2.6\cdot10^{-17}$ & 260 \\
NA-B150A2 & 12 & 0.2  & $3.1\cdot10^8$ & $3.8\cdot10^{-17}$ & 220 \\
NA-B150B1 & 16 & 0.5  & $4.4\cdot10^8$ & $3.6\cdot10^{-17}$ & 360 \\
NA-B150B2 & 11 &$<0.1$& $3.6\cdot10^8$ & $3.1\cdot10^{-17}$ & 160 \\
NA-B90A   & 12 & 0.8  & $4.7\cdot10^8$ & $2.6\cdot10^{-17}$ & 210 \\
NA-B90B   & 14 & 0.8  & $3.5\cdot10^8$ & $4.0\cdot10^{-17}$ & 240\\
\tableline
\end{tabular}
\end{center}
\caption{Pre-flight voltage noise measurements}
\label{tab:noise}
\end{table}

\item Spectral characterization was made using a Lamellar Grating
Interferometer with a Hg~Vapor Lamp as Rayleigh-Jeans source. 
A check for near-band leak in the optical filters 
was made by means of three thick grill filters at 4.9, 7.7 and 
10.5~cm$^{-1}$, which reduced the integrated signal by more than a 
factor 100 in the corresponding bands.

\item Beam profiles ware measured using a collimated 
source, filling the telescope acceptance area. 
The wide (1 meter diameter) parallel beam is produced
by a thermal source in the focus of a large parabolic reflector.

\item Sidelobes measurements or upper limits. 
We have illuminated the fully integrated payload using a high power (30~mW) 
90~GHz Gunn oscillator, completed with calibrated attenuators and 
high gain horn (20~deg FWHM). The microwave source was electronically 
chopped at 11 Hz. The payload was located at the center of a wide, flat
area, and the source was setup at a distance of $\sim50$~m from the payload, 
at an apparent elevation of $\sim39$~degrees, so that the microwave beam 
over-illuminated the payload. We first boresighted the \boom 
telescope to the fully attenuated microwave source, to record the
axial gain of the instrument. We then rotated the payload in azimuth, 
reducing the source attenuation as necessary, to record the far 
sidelobe response of the instrument for different off-axis locations 
of the source. We made several spins, with different apparent
elevations of the source. Results for a sample elevation are 
reported in Figure~\ref{fig:slobes}.
\end{enumerate}

\section{Observations and in flight performances}
\label{sec:obs}

\begin{figure}[ht]
\epsscale{.6}
\begin{center}
\plotone{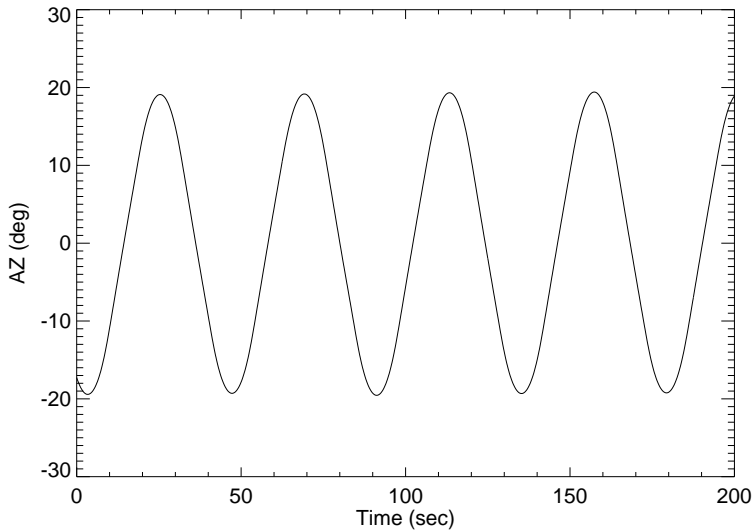}
\caption{
Azimuth read by the magnetometer during a scan. The amplitude is
$\pm 20$~deg, the shape is a smoothed saw tooth function.
In this configuration the fraction of the scan at
constant speed fraction is about one half of the total.
\label{fig:scan}} 
\end{center}
\end{figure}

The system was flown for 6 hours on 1997 August 30 from the
National Scientific Balloon Facility in Palestine, Texas.
All the subsystems performed well during the flight:
the He vent valve was opened at float and was closed at termination,
the Nitrogen bath was pressurized to 1000 mbar,
the $^3$He fridge temperature (290mK) drifted with 
the $^4$He temperature by less than 10~mK during the 7.5~hours 
of the flight, with a maximum temperature of 300~mK before venting 
to atmosphere and slowly drifting down to 290~mK.
The main $^4$He cryostat wormed to 2.05~K during ascent and 
then recovered to 1.65~K.
Pendulations were not generated during CMB scans at a
level greater than 0.5 arcmin, and both azimuth scans 
at 2~deg~s$^{-1}$ (see Figure~\ref{fig:scan})
and full azimuth rotations (at 2 and 3 rpm) 
of the payload were performed effectively.
The loading on the bolometers was as expected,
and the bolometers were effectively CR immune, with
white noise ranging between 500 and 1000 $\mu {\rm K} /\sqrt{{\rm Hz}}$. 

\begin{figure}[ht]
\epsscale{.6}
\begin{center} 
\plotone{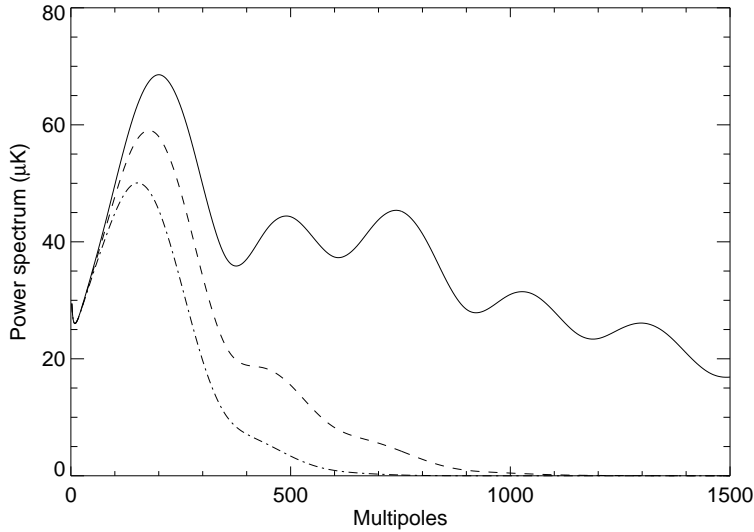}
\caption{
Convolution of the best fit power spectrum of CMB
anisotropy \cite{b97mausk} with the BNA beams.
The continuum line is the original power spectrum, the dashed line
is the convolution with a 150~GHz channel(16.6 arcmin FWHM) and the
dash-dotted line is the convolution with a 90~GHz channel
(26.0 arcmin FWHM).
The importance of a good determination of the beam shape is evident.
\label{fig:laura}}
\end{center}
\end{figure}

Measurements of the in flight performance (time constants,
beam mapping and calibration constants) are essential to a
good determination of the power spectrum of the anisotropies.
The measured power spectrum is in fact the convolution of the real 
power spectrum with an angular function given by
the shape of the beam (see Figure~\ref{fig:laura}) and
the amplitude of the spectrum depends on the calibration
constant.

\subsection{In-flight time constants of the bolometers}
\label{time_c}

The total transfer function of the system is the combination
of the electronics and the bolometer transfer functions.
A high pass filter ($\tau \sim 10$~s) and a two pole
Butterworth low pass filter determine the electronic transfer
function, which was measured in lab.
From equation (\ref{eqn:eqbolo}) we see that the bolometer behaves like
a low pass filter with a time constant $\tau=C/G$. 
The value of $\tau$ changes with the temperature of the receiver 
and with the radiative input, and has to be measured in flight.
For a given input ($IN(t)$) on the bolometer, the output ($OUT(t)$)
of the system with transfer function $T(\omega)$ (bolometer and 
electronics) is 
\begin{equation}\label{eqn:fft}
OUT(t)=IFT(FT(IN(t))\cdot T(\omega))
\end{equation}
where $FT$ and $IFT$ are the Fourier Transform and 
Inverse Fourier Transform operator respectively.
It is possible to fit the parameters in the $T(\omega)$
using the $OUT(t)$ data. As input we use the signal
from a planet in a fast scan mode. When the scan speed is 3 rpm
(18~deg~s$^{-1}$) the time of transit of the planet in the beam 
$( < 40^\prime)$ is about the time of two samples 
(sampling rate is 62.5~Hz).
So the $IN(t)$ signal has to be modeled with a beam shape
function. Simulations show that for this sample rate and scan speed the
results are the same within the error, assuming either 
gaussian or square beam. 
The measured time constants are reported in Table \ref{tab:timec}.
These values are longer than expected for spider web bolometers.
The source of the problem was tracked to excess heat capacity
of the chromium layer, and was corrected for the devices used
in the subsequent LDB flight.

\begin{table}[ht]
\begin{center}
\begin{tabular}{lc} \tableline \tableline
channel            &Time constant (ms)\\ \tableline
NA-B150A1          &  $102 \pm 11$    \\
NA-B150B1          &  $ 83 \pm 13$    \\
NA-B150B2          &  $ 83 \pm 12$    \\
NA-B90A            &  $165 \pm  9$    \\
NA-B90B            &  $ 71 \pm  8$    \\ \tableline
\end{tabular}
\end{center}
\caption{
In-flight time constants of the \boom NA bolometers. All values 
except NA-B150B2 are calculated from fast scans across Jupiter.
The NA-B150B2 value is calculated by comparing that detector's 
cosmic ray signals with those of NA-150B1, which are the same
to within error.}
\label{tab:timec}
\end{table}

\subsection{Jupiter calibration and beam mapping}
\label{jupitercal}

During the August, 1997 North American flight of \boomn, we
calibrated the instrument and produced a detailed beam map by scanning
the planet Jupiter \cite{jupiter}. 
In addition, we made a secondary calibration 
through measurements of the CMB dipole, which is known to $\simeq 1$\% 
accuracy from the measurements of the COBE satellite \cite{cobedipole}.

The responsivity of the instrument is defined as 
\begin{equation}
{\mathcal{R}} = {\Delta V \over \Delta W}
\end{equation}
Where $\Delta W$ is the radiative input and $\Delta V$ the output of 
the bolometers. The calibration constant directly converts the signal 
in Volt into CMB temperature (in Kelvin) and is defined as
\begin{equation}
{\mathcal{K}} = {\Delta V \over \Delta T_{CMB}}
\end{equation}

The use of planets for calibration is a standard in mid latitude CMB 
experiments. Planets are bright sources, and their mm-wave brightness 
temperature is known at the 5\% level. Moreover, they are point-like 
sources when compared to our beam, 
so they are perfectly suitable for mapping the shape of the beam pattern 
of the telescope.

The signal from the planet is
\begin{equation}\label{eqn:v_planet}
\Delta V_{planet}={\mathcal{R}} A \Omega_{planet} \int E(\nu)
BB(T_{eff},\nu)
d\nu         
\end{equation}
where $T_{eff }$ is the brightness temperature of the planet, 
$E(\nu)$ is the spectral efficiency and $\Omega_{planet}$ is 
the (small) solid angle filled by the planet
as computed from the ephemerides on the day of 
the observation.  The CMB signal is related to the
derivative of the Planck 
function with respect to the temperature:
\begin{equation}\label{eqn:v_cmb}
\Delta V_{CMB}= {\mathcal{R}} A \Omega {\Delta T_{CMB} \over T_{CMB}}
\int E(\nu) BB(T_{CMB},\nu) {x e^x \over e^x-1}   d\nu
\end{equation}
where $A \Omega$ is the throughput of the system, 
$x = h\nu / k T_{CMB}$ and $BB(\nu, T)$ is the Planck function.
Note that the beam solid angle $\Omega$ is the angular response function 
$RA(\theta,\Phi)$ integrated over all the angles :
\begin{equation}
\Omega=\int_{4 \pi}  RA(\theta,\Phi)  \cos \theta  \sin \theta d\theta
d\Phi
\end{equation}
From the (\ref{eqn:v_planet}) and the (\ref{eqn:v_cmb})
the resulting expression for the calibration constant is:
\begin{equation}
{\mathcal{K}}={ \Delta V_{CMB} \over \Delta T_{CMB}} = 
{ \Delta V_{planet} \over T_{CMB} } 
{ \Omega \over \Omega_{planet} }
{\int E(\nu) BB(T_{CMB},\nu) x e^x  (e^x-1)^{-1} d\nu
\over 
\int E(\nu) BB(T_{eff},\nu) d\nu }
\end{equation}
$E(\nu)$ has been measured in laboratory, 
$\Delta V_{planet}$ is the maximum signal from the planet and
$\Omega$ can be determined from a raster scan on the
planet.

\begin{figure}[ht]
\epsscale{.6}
\begin{center} 
\plotone{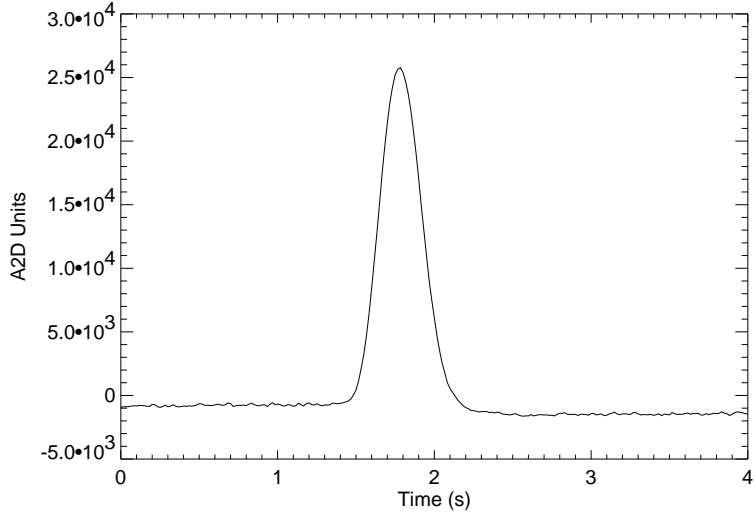}
\caption{
Raw data of a single scan on Jupiter at 150~GHz.
S/N is $>100$, scan speed is 1.1~deg~s$^{-1}$.
The negative tail of the data is due to the AC-coupling
of the signal.
\label{fig:jup}}
\end{center}
\end{figure}

\begin{figure}[ht]
\epsscale{.45}
\begin{center}
\rotatebox{90}{
\plotone{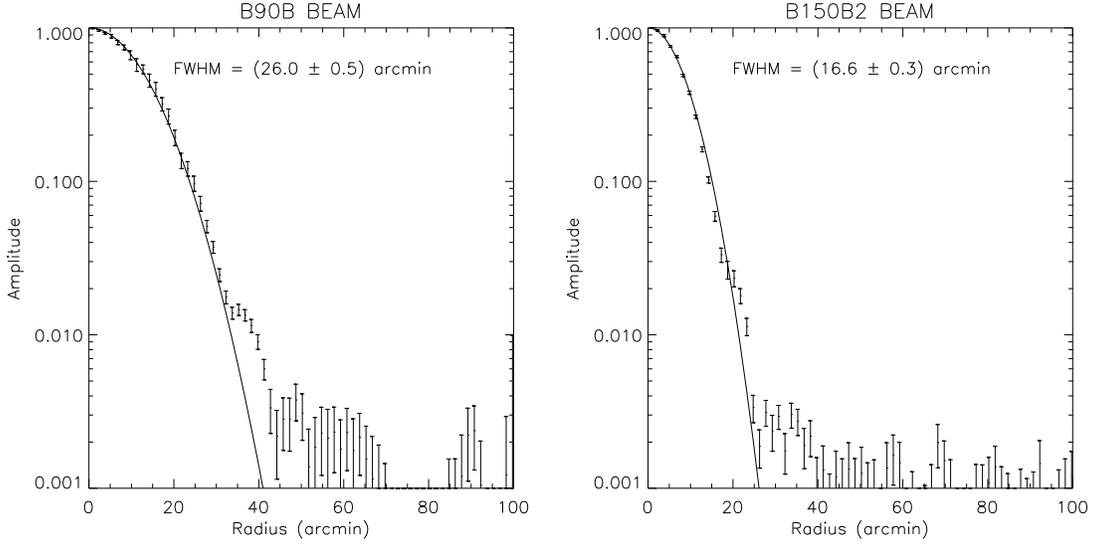}}
\caption{
Jupiter data averaged in annuli around the centroid and gaussian fits.
The small shoulders evident at a few percent level are real and are
due to aberrations in the optical system.
\label{fig:beam}}
\end{center}
\end{figure}

The telescope made two series of scans on Jupiter with an amplitude
of 15~deg and a speed of 1.1~deg~s$^{-1}$.
Jupiter has an effective source temperature of $T_{eff}=(173 \pm 9)$~K 
\cite{Ulich,MSAM}.
The data have a high signal to noise ratio ($> 100$)
(see Figure~\ref{fig:jup})
and permit to produce the beam profile of each receiver.
The solid angles $\Omega$ are computed integrating the 
angular response, $RA(\theta,\Phi)$, given by the pixellized
Jupiter map. Errors are given by the pointing error and the
noise. The beams are symmetric with minor and major axes
equivalent within 5\%.
FWHM are computed averaging the data in annuli around 
the center and fitting the shape with a gaussian function
(see Figure~\ref{fig:beam}).

Solid angles and FWHM are summarized in Table~\ref{tab:omega}.
The calibration constants are in Table \ref{tab:cal}.

\begin{table}[ht]
\begin{center}
\begin{tabular}{lrc} 
\tableline 
\tableline
channel     &     FWHM (arcmin)& $\Omega(sr)$         \\ 
\tableline
NA-B150A1          &    19.5   & $(3.04 \pm 0.17)10^{-5}$   \\
NA-B150B1          &    19     & $(3.03 \pm 0.25)10^{-5}$   \\
NA-B150B2          &    16.6   & $(2.63 \pm 0.10)10^{-5}$   \\
NA-B90A            &    24     & $(5.61 \pm 0.36)10^{-5}$   \\
NA-B90B            &    26     & $(6.47 \pm 0.27)10^{-5}$   \\ 
\tableline
\end{tabular}
\end{center}
\caption{Beam size measurements for the \boom NA telescope.
}
\label{tab:omega}
\end{table}

\begin{table}[ht]
\begin{center}
\begin{tabular}{lcc} \tableline \tableline
channel& ${\mathcal{K}}_{Jupiter}(nV/mK)$&${\mathcal{K}}_{dipole}(nV/mK)$\\ \tableline
NA-B150A1          &  $57.1 \pm 5.1$    &     $54.1 \pm 3.2$         \\
NA-B150B1          &  $59.9 \pm 7.8$    &     $53.0 \pm 2.1$         \\
NA-B150B2          &  $71.4 \pm 5.7$    &     $66.4 \pm 1.5$         \\
NA-B90A            &  $48.1 \pm 8.2$    &     $43.1 \pm 1.1$         \\
NA-B90B            &  $60.9 \pm 8.5$    &     $61.9 \pm 2.5$         \\ \tableline
\end{tabular}
\end{center}
\caption{Calibration constants for the \boom NA telescope.
The dipole calibration comes from the set of revolutions
immediately after the Jupiter calibration.}
\label{tab:cal}
\end{table}

\subsection{Dipole calibration}
\label{dipolecal}

The dipole \cite{dipole} is a well calibrated source, 
featuring the same spectrum 
as CMB anisotropies and completely filling the beam of CMB anisotropy 
experiments. It is available at any time during the observations, thus 
allowing repeated checks for the calibration of the experiment. If a 
scanning experiment can perform large scale scans, the dipole will appear 
as a scan synchronous signal with an amplitude $\Delta T_{dipole}$ of the 
order of a few mK (depending on the actual scan geometry), thus
perfectly suitable for calibration. The calibration constant is simply
\begin{equation}
{\mathcal{K}} = {\Delta V_{dipole} \over \Delta T_{dipole}}
\end{equation}
where $\Delta V_{dipole}$ is the amplitude of the signal during
dipole observations. 
The precision of this measurement is affected by the presence
of atmospheric emission and by 1/f noise of the system.
The BNA bolometers showed noise spectra white down to~10 mHz. In these
conditions it is possible to make dipole scans with a period of
the order of a minute without being significantly affected by 1/f
noise.

\begin{figure}[ht]
\epsscale{.6}
\begin{center}
\plotone{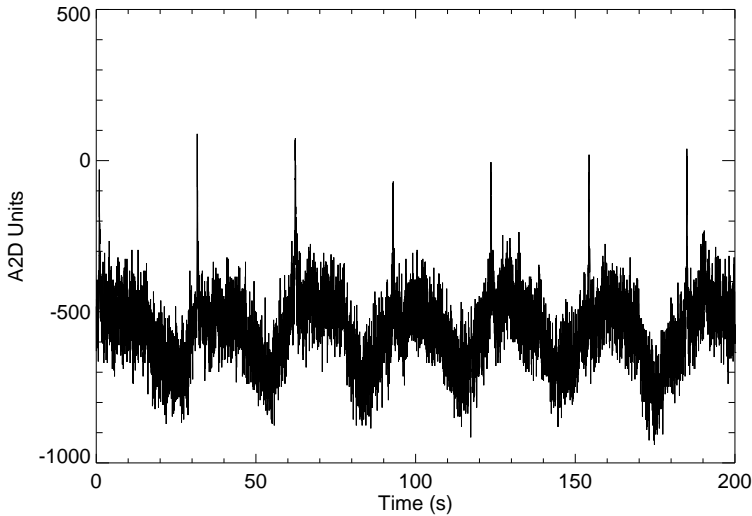}
\caption{
Raw data of a fast rotation. The CMB dipole produces
a sinusoidal signal. The spike at regular phase is the
signal from Jupiter used for time constant measurements.
Full rotations of the payload have been repeated 4 times
during the flight, at intervals of more than one hour.
\label{fig:dipole}}
\end{center}
\end{figure}

Dipole scans consisted in full azimuth revolutions of the 
gondola at 18~deg~s$^{-1}$ and were carried out for 10 minutes every hour.  
The instantaneous signal to noise ratio of the dipole scans was $\sim$3.  
Raw data from one of the dipole
scans are shown in Figure~\ref{fig:dipole}.

\begin{figure}[ht]
\epsscale{.6}
\begin{center}
\rotatebox{270}{
\plotone{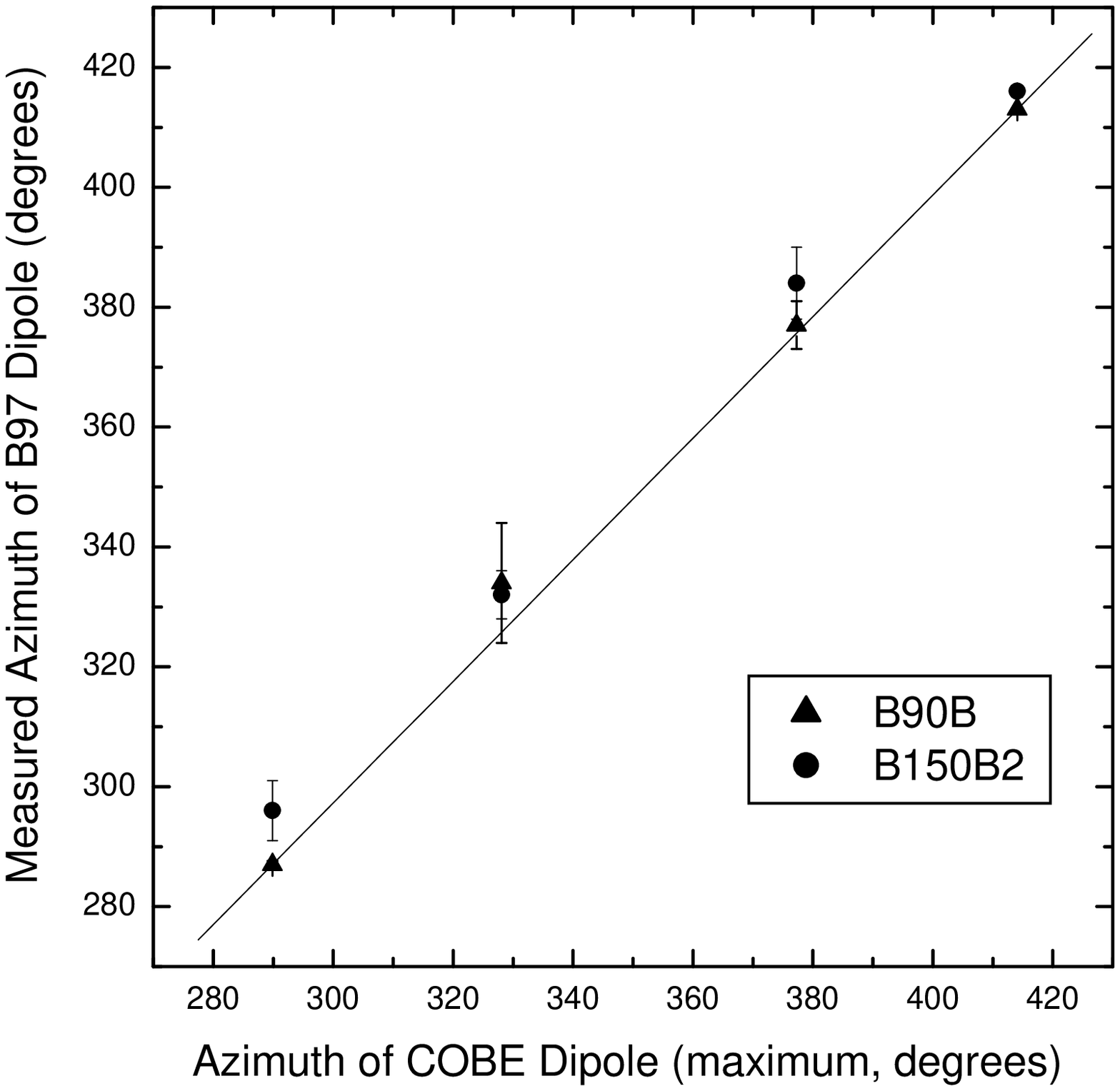}}
\caption{
Azimuth of the maximum signal versus azimuth of
the expected dipole direction during fast rotations.
The good correlation is a proof that the
signal is originated by the CMB dipole
and it is not a local effect.
\label{fig:dipoloaz}}
\end{center}
\end{figure}

In Figure~\ref{fig:dipoloaz} we plot the azimuth of the maximum dipole
signal versus the azimuth of the CMB dipole as observed from the 
payload position at the time of the observations, thus providing evidence
for detection of the CMB dipole, and convincingly rejecting 
any hypothesized local origin of the signal.

\begin{figure}[ht]
\epsscale{.6}
\begin{center} 
\rotatebox{270}{
\plotone{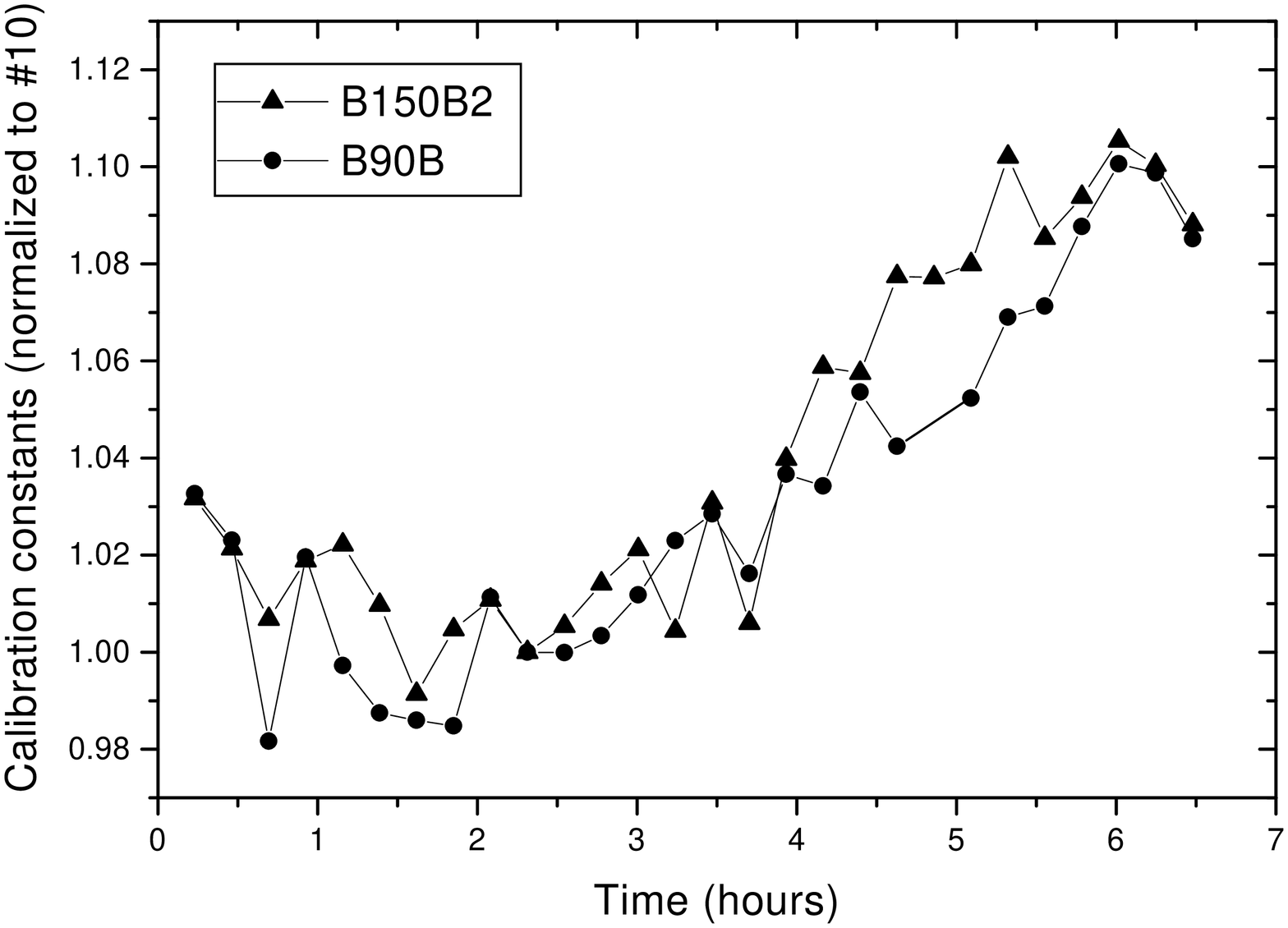}}
\caption{
Fractional responsivity variations during the flight
as measured from the internal
calibrator signals, for both the channels used in the analysis.
Starting from the second hour of flight the variations are
less than~10\%. The first two hours aren't used in the data analysis.
\label{fig:callamps}}
\end{center}
\end{figure}

The CMB dipole calibration in the first set of revolutions is consistent
with Jupiter calibration performed immediately before 
(see Table \ref{tab:cal}). We note however that the dipole 
signal shape and size changes in the subsequent set of rotations
for both B90B and B150B2 channels, producing calibrations 
not completely consistent with those derived from the internal calibrator
(see Figure~\ref{fig:callamps}).

Even if the atmospheric emission is greatly reduced at balloon altitude
($ > 35$~km), large scale fluctuations of the atmospheric brightness 
could be non-negligible with respect to the CMB dipole, thus giving 
a contamination at large scales hard to remove without a
monitoring high frequency (240, 400~GHz) channel \cite{Page,max3k}. 
For this reason we have used the Jupiter calibration and the
internal calibrator transfer for the data analysis of this flight 
of \boomn.

\subsection{In flight noise}
\label{noise}

\begin{figure}[ht]
\epsscale{.6}
\begin{center}
\plotone{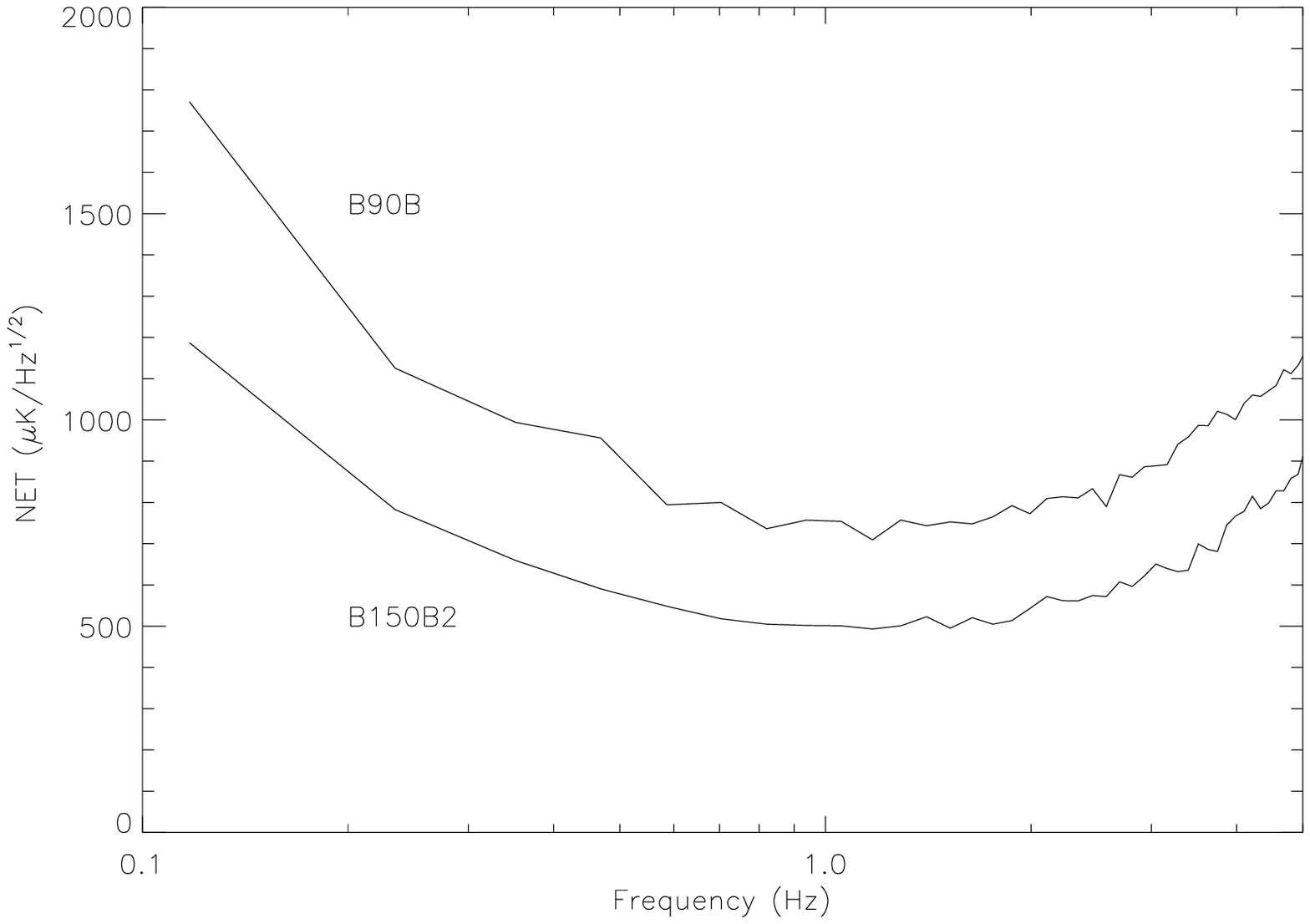}
\caption{
Noise Equivalent Temperature versus frequency for the two channels
used in the analysis of the angular power spectrum of the CMB.
At low frequencies 1/f~noise dominates. The rise at high
frequencies is due to the bolometer time constants: the receiver
looses sensitivity and the NET is increased after deconvolution.
The most interesting frequencies are around 1.5~Hz where is expected
to be the first peak of the anisotropies in the CMB (with scan
speed of 2~deg~s$^{-1}$ at the elevation of 41~deg).
\label{fig:noise}}
\end{center}
\end{figure}

In flight noise performance is measured computing the power
spectrum of the bolometers signals after de-spiking and deconvolution,
avoiding data taken when the gondola inverts the scanning
direction (turnarounds).
Noise equivalent temperatures (NET) are calculated using in
flight calibration constants. In flight noise includes
bolometer noise, electronics noise, signal from the atmosphere,
radio frequency noise, signal from the Galaxy and signal 
from the CMB.
Results are in Table~\ref{tab:fnoise}. Power spectra are reported
in Figure~\ref{fig:noise} for the channels used in the analysis.

\begin{table}[hb]
\begin{center}
\begin{tabular}{lc} \tableline \tableline
Channel         &  NET$_{CMB}~(\mu K/\sqrt{Hz})$ \\ \tableline
NA-B150A1       &  700   \\
NA-B150B1       &  1000  \\
NA-B150B2       &  500   \\
NA-B90A         &  1000  \\
NA-B90B         &  740   \\ \tableline
\end{tabular}
\end{center}
\caption{Noise Equivalent Temperature measured in flight.}
\label{tab:fnoise}
\end{table}

\section{Sensitivity and Data Analysis}
\label{sens}

The sensitivity of a receiver to fluctuations in sky brightness at
different angular scales can be represented by a window function.  The
shape of the window function depends on the details of the measurement.
Calculations have been made for the window functions of different
experiments (e.g. 
\cite{WSS}) taking into account different beam shapes
and sizes, chopping strategies, signal processing electronics, and
data analysis strategies.  Because of the complex chopping strategies
employed to obtain stable offsets in these experiments, the window
functions are often complicated.  Recent experiments attempt to
use specially tailored filters to generate multiple well-defined
window functions from a single chop (Big Plate, MSAM).
In addition, upcoming space missions will attempt to make fully
reconstructed maps of diffuse millimeter-wave emission (MAP,
Plank).  For all of these experiments, the scan strategy
is driven by the achievable bandwidth of the detectors which limits
the maximum and minimum scan speed.

In \boom we have designed our scan strategy and beam size to
utilize the full bandwidth and sensitivity of our detectors.
The scan speed for both flights of \boom is matched
to the signal bandwidth of the detectors and the sensitivity to
large angular scales is limited by the stability of the readout
electronics and atmospheric fluctuations.  \boom can operate
in either total power mode with the data from each detector
analyzed independently, or in chopped mode, differencing between
symmetric pixels in the focal plane to remove common mode noise
sources.  In total power mode, the window function of each channel
is limited only by the beam size and the scan length.  In differenced
mode, the window function is determined by the beam size and
beam separation in the focal plane. 

The data from the BNA flight have been
analyzed as a pixellized map with an experimentally determined
correlation matrix describing the noise. 

\begin{figure}[ht]
\epsscale{.8}
\begin{center}
\plotone{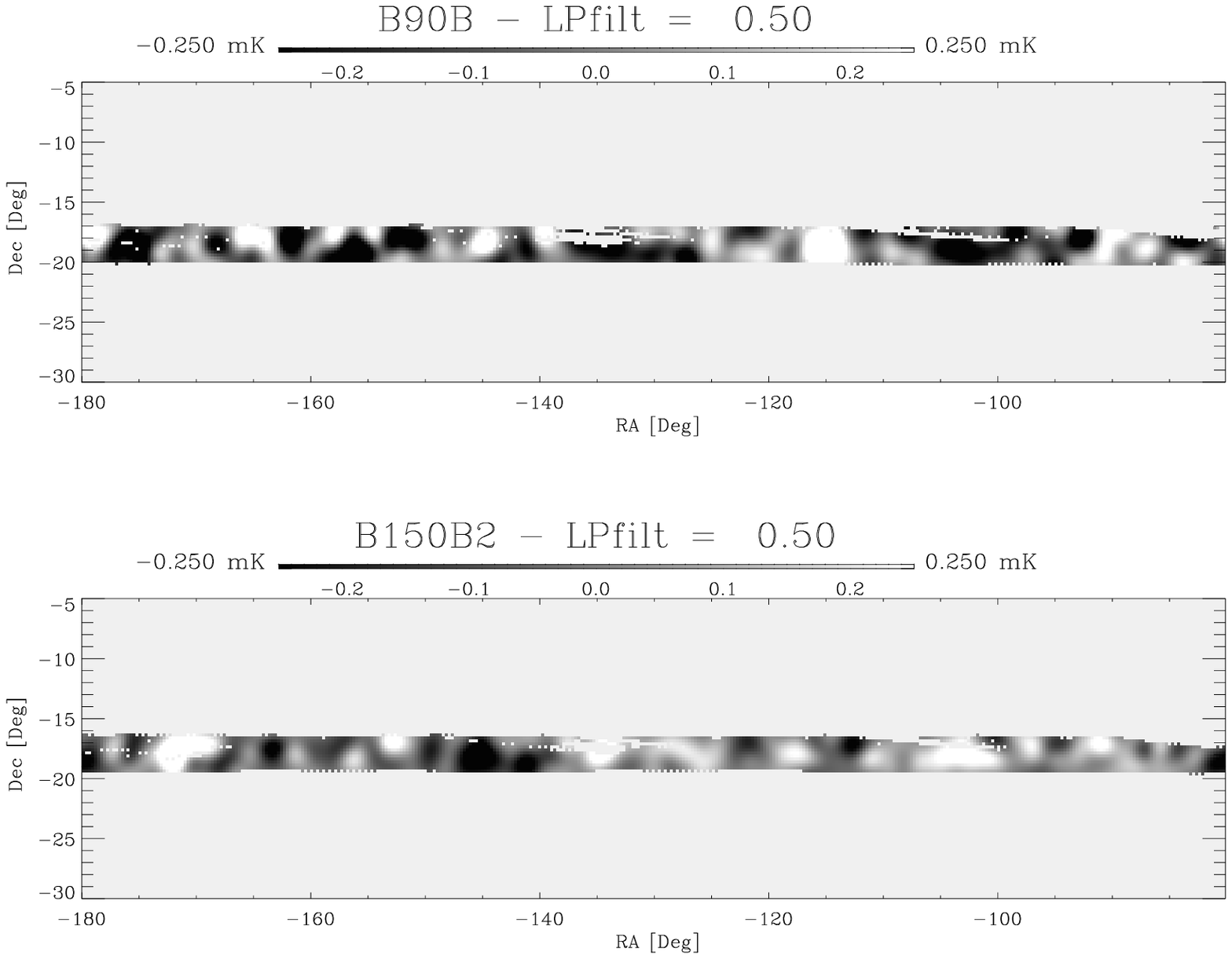}
\caption{
Maps of the measured flight from B150B2 and the B90B channels.
Both maps are smoothed
with a 0.5~deg FWHM gaussian in order to increase the
signal to noise ratio. The structures in the maps are only partially
concordant,
showing that these maps are noise dominated rather than signal dominated.
For better visualization only a portion of the full map (see coverage
in Figure~\ref{fig:coverage}) is reported here.
\label{fig:maps}} 
\end{center}
\end{figure}

Maps of the B150B2 and B90B channels are in Figure~\ref{fig:maps}.
The $\sim 25000$~pixels maps are noise dominated. 
These maps are produced just filtering the Time Ordered Data and
coadding the data in pixel with the HEALPix 
scheme\footnote{http://www.eso.org/kgorski/healpix/}. 
The map produced
to compute the anisotropies power spectrum was made using the MADCAP
software 
package\footnote{http://cfpa.berkeley.edu/$\sim$borrill/cmb/madcap.html}.
The noise correlation function is estimated from the 
time ordered data assumed to be noise dominated.

This analysis, described in \cite{b97mausk}, lead to an
error on the power spectrum measurements of $\sim \pm 15 \mu$K
at 68\% confidence level for band powers averaged over 
$\Delta \ell = 50$ bins, centered at 6 multipoles covering the range
$25 < \ell < 325$.

\section{Conclusions}
\label{conc}

The feasibility of extended (many hundred square degree), resolved
(20$^\prime$~FWHM), sensitive CMB maps using quasi-total-power 
balloon-borne microwave photometers has been demonstrated 
with the \boomn/NA instrument. Critical technologies have
been developed in several areas: spider web bolometers, total
power readout electronics, low sidelobe response telescope, long duration 
cryogenics, and a scan-oriented attitude control system. 

The \boomn/LDB \cite{crill} payload makes optimal use of
the technologies described here providing the first high
signal to noise ratio map \cite{nature} of a wide portion of the 
microwave sky.

\section{Acknowledgements}
\label{sec:akn}
The \boom experiment has been supported by Programma
Nazionale di Ricerche in Antartide, Universit\'a di Roma
``La Sapienza'', and Agenzia Spaziale Italiana in
Italy, by NSF and NASA in the USA, and by PPARC in the
UK. Doe/NERSC provided the supercomputing facilities.
We acknowledge the use of HEALPix.

\clearpage
\newpage
\markright{REFERENCES}

\end{document}